\documentclass[preprints,article,accept,moreauthors,LaTeX,10pt,a4paper]{mdpi}

\firstpage{1}
\articlenumber{x}
\doinum{10.3390/------}
\pubvolume{xx}
\pubyear{2017}
\copyrightyear{2017}
\history{~}
\pdfoutput=1

\usepackage[english]{babel}
\usepackage[mathscr]{eucal}
\usepackage{bm,courier,wrapfig,subcaption,wasysym,amsbsy,amsfonts,amssymb}

\def\t{\mathfrak{t}}
\def\tt{\tau}
\def\ww{w}
\def\vv{v}
\def\zz{y}
\def\rr{\rho}
\def\x{\textbf{x}}
\def\y{\textbf{y}}
\def\q{\textbf{q}}
\def\p{\textbf{p}}
\def\del{\delta}
\def\de{\partial}
\def\lap{\Delta}
\def\Asc{\mathscr{A}}
\def\BB{\mathscr{B}}
\def\FF{\mathscr{F}}
\def\G{\mathscr{G}}
\def\LL{\mathcal{L}}
\def\KM{\mathfrak{K}}
\def\Eu{\mbox{E}_1}
\def\EE{\mathscr{E}}
\def\at{\lambda}
\def\ee{\eps}
\def\Dom{\mbox{Dom}}
\def\AA{\mathcal{A}}
\def\AAa{\mathcal{A}}
\def\AAae{\mathcal{A}_{\ee}}
\def\u{u}
\def\mm{\kappa}
\def\gEM{\gamma_{EM}}
\def\be{\beta}
\def\ga{\gamma}
\def\eps{\varepsilon}
\def\lam{\varrho}
\def\si{\sigma}
\def\te{\theta}
\def\vfi{\varphi}
\def\Om{\Omega}
\def\Ga{\Gamma}
\def\N{\mathbb{N}}
\def\R{\mathbb{R}}
\def\C{\mathbb{C}}
\def\Fock{\mathfrak{F}}
\def\Fi{\hat{\phi}}

\def\Fiue{\Fi^{\u}_{\epsilon}}
\def\Ti{\hat{T}}

\def\Tiue{\Ti^{\u,\ee}}
\def\Tc{T^{(\lozenge)}}
\def\Tnc{T^{(\blacksquare)}}
\def\rist{\upharpoonright}
\def\geqs{\geqslant}
\def\l{\left}
\def\r{\right}
\def\la{\langle}
\def\ra{\rangle}
\def\dd{\displaystyle}
\def\Im{\mbox{Im}\,}
\def\Re{\mbox{Re}\,}
\def\b0{\mathbf{0}}
\def\parn{\par}

\Title{Local Casimir Effect for a Scalar Field in Presence of a Point Impurity}

\Author{Davide Fermi $^{1,2,\dagger}$, Livio Pizzocchero $^{1,2\ddagger}$}

\AuthorNames{Davide Fermi, Livio Pizzocchero}

\address{%
$^{1}$ \quad Dipartimento di Matematica, Universit\`a di Milano, Via C. Saldini 50, I-20133 Milano, Italy \\
$^{2}$ \quad Istituto Nazionale di Fisica Nucleare, Sezione di Milano, Italy \\
$^{\dagger}$ \quad davide.fermi@unimi.it \\
$^{\ddagger}$ \quad livio.pizzocchero@unimi.it}

\abstract{The Casimir effect for a scalar field in presence of delta-type potentials
has been investigated for a long time in the case of surface delta functions,
modelling semi-transparent boundaries. More recently Albeverio,
Cacciapuoti, Cognola, Spreafico and Zerbini \cite{AlbCog,AlbCac,SprZer} have considered
some configurations involving delta-type potentials concentrated at points
of $\R^3$; in particular, the case with an isolated point singularity
at the origin can be formulated as a field theory on $\R^3\!\setminus\!\{\b0\}$,
with self-adjoint boundary conditions at the origin for the Laplacian.
However, the above authors have discussed only global aspects of the Casimir effect,
focusing their attention on the vacuum expectation value (VEV) of the total energy.
In the present paper we analyze the local Casimir effect with a point delta-type
potential, computing the renormalized VEV of the stress-energy tensor at any point of
$\R^3\!\setminus\!\{\b0\}$; to this purpose we follow the zeta regularization
approach, in the formulation already employed for different configurations in
previous works of ours (see \cite{FP3,FP4,FPBook} and references therein).}



\begin{document}

\section{Introduction}
The main characters in investigations on vacuum effects of the Casimir type
are the boundary conditions assumed for the quantum field and/or the external
potential possibly acting on the field itself
\cite{ActHarm,BorNew,BorNew2,Ful1,Brev,Casimir,ptp,FP3,FP4,Ful2,Mil,MilSoft,BrevSym,Ful3,Ful4}.
The boundary conditions are typically employed to account for the presence of perfectly
conducting walls, or perfectly reflecting mirrors. On the other hand, the interpretation
of the external potentials depends essentially on their structure; in many cases
these potentials can be viewed as modelling some type of confinement, softer than
the one given by sharp boundaries.
\parn
Nowadays a quite remarkable literature is available, regarding Casimir-type
settings with external delta-type potentials. Such models can be viewed as limit
cases of configurations with sharply peaked (but regular) potentials.
\parn
Most of the literature considers the case of surface delta functions,
concentrated on supports of co-dimension $1$; these are commonly interpreted
as semi-transparent walls, inducing a partial confinement of the quantum field.
The first ones to investigate a Casimir configuration of this kind were probably
Mamaev and Trunov \cite{MaTru}, who computed the renormalized VEV of the energy
density for a massive scalar field in presence of delta potentials concentrated
on two parallel plates.
Variations of the same model, concerning both a scalar and a spinor field, were later
examined by Bordag, Hennig and Robaschik in \cite{Bord}.
In the last two decades, there has been a renewed interest on surface delta potentials:
see, e.g., \cite{BBK,Bra,CFP,DelCil,MilBook,Graham,Khus,Milt,Cast,Scan}.
\parn
The Casimir effect in presence of point delta-type potentials (concentrated on
supports of co-dimension $3$) has been studied only in more recent times, and the
existing literature is not so wide; these configurations are typically interpreted
in terms of point-like impurities.
In \cite{SprZer}, Spreafico and Zerbini proposed a general setting to renormalize
the relative partition function of a finite-temperature quantum field theory
(on flat or even on curved, ultrastatic spacetimes with noncompact spatial section);
in this work the authors discussed, as an application, the total Casimir energy
at finite or zero temperature for a massless scalar field (in flat spacetime),
in presence of one or two point-like impurities. In \cite{AlbCog} Albeverio, Cognola,
Spreafico and Zerbini computed the renormalized, relative partition function and the
Casimir force for a massless scalar field in presence of an infinite conducting plate
and of a point-like impurity, placed outside the plate. A similar analysis was
performed in \cite{AlbCac} by Albeverio, Cacciapuoti and Spreafico, who determined
the renormalized, relative partition function for a massless scalar field in presence
of a point-like impurity and of a Coulomb potential centered at the same point.
\vspace{0.1cm} \parn
From a mathematical point of view, the description of delta-type potentials can be
given in terms of suitable boundary conditions across the support of the delta functions,
defining a self-adjoint realization of the Laplace operator. This approach has been
developed for delta functions concentrated on surfaces, curves or points in $\R^3$.
So, a problem involving $-\lap$ (the opposite of the Laplacian) plus a delta-type
potential is reformulated with full analytical rigor as a problem in the region outside
the singularity, where the fundamental operator is $-\lap$ with the above mentioned
boundary conditions.
When this setting was originally devised, the interest in delta-type potentials was
not motivated by their action on quantum fields but, rather, by non-relativistic
quantum mechanics; the aim was to define rigorously Schr\"odinger operators with
delta-type potentials and to develop, in particular, the corresponding scattering theory.
Of course, the operator $-\lap$ plus a delta-type potential has a different status
in quantum field theory, where it can appear in the spatial part of the field equations.
\parn
In the case of a point delta-type potential on $\R^3$, the rigorous definition of the
corresponding operator in terms of boundary conditions for the Laplacian was first given
in a seminal paper of Berezin and Faddeev \cite{BeFa}; a standard reference on this topic,
using systematically the language of Sobolev spaces, is the book by Albeverio \textsl{et al.} \cite{AlbB}.
To implement this setting, a price must be paid: one must think the potential as the product
of the point delta function by an infinitesimally small coupling constant.
It is customary to interpret the infinitesimal nature of this constant as the effect
of some ``renormalization'' of the interaction, an idea suggested by the construction of \cite{BeFa}.
\parn
Before proceeding, let us mention that general delta-type potentials have also been treated
within the framework of much more general mathematical theories; in particular, they have
been described in terms of singular perturbations of self-adjoint operators in scales of
Hilbert spaces by Albeverio \textsl{et al.} \cite{Alb2,Alb3} and by means of Krein-like
resolvent formulas by Posilicano \textsl{et al.} \cite{PosMan,PosKre}.
\parn
In the present work we analyze the Casimir physics of a massless scalar field in presence of
a point-like impurity. This configuration is closely related to the settings of \cite{AlbCog,AlbCac,SprZer};
however these papers discussed only global observables, like the total energy. On the contrary,
our analysis is focused on local aspects; more precisely, we compute the renormalized VEV
of the stress-energy tensor at any space point outside the impurity.
To treat the point delta-type potential, we stick to the standard setting of \cite{AlbB,BeFa}.
Besides, to renormalize the stress-energy VEV we follow the \textsl{local zeta regularization}
approach; here, a regularization is introduced for the field theory depending an a complex parameter,
and the renormalization of local observables is defined in terms of the analytic continuation
with respect to this parameter.
\parn
Zeta regularization is an elegant strategy to give meaning to the divergent expressions
appearing in \textsl{na\"ive} manipulations of quantum field theory; its application to
the local observables of quantum fields was proposed by Dowker and Critchley \cite{DowKri},
Hawking \cite{Hawk} and Wald \cite{Wal}, and especially supported by
Actor, Cognola, Dowker, Elizalde, Moretti, Zerbini \textsl{et al.},
who must be credited with developing this idea in a systematic way
(see \cite{ActBox1,Actor,DowKen,MorBo} and the references cited therein).
The same ideas have become more popular in the treatment of global observables (such as the
total energy), resulting into an abundant literature (see, e.g., \cite{Eli1,Eli2,Kir1}
and references therein); notably, global zeta regularization appears in all
the previously cited works on field theory with point-like singularities.
In our recent book \cite{FPBook}, we have proposed a formulation of the local (and global)
zeta techniques for a scalar field, based on canonical quantization and on the introduction
of a suitably regularized field operator depending on a parameter $\u \in \C$;
the renormalization of local (or global) observables is defined in terms of the analytic
continuation to a neighborhood of the point $\u=0$, formally corresponding to the
unregularized field operator.
\parn
From the very beginning of zeta regularization theory, it was understood that the analytic
continuation required by this approach is deeply related to certain integral kernels
associated to the fundamental operator of the field theory, i.e., $-\lap$ plus the
possibly given external potential. Here we mention, in particular, the Dirichlet
and heat kernels which correspond, respectively, to the complex powers and to the exponential
of the fundamental operator; these facts are relevant even for the results described in
the present paper.
\parn
Let us describe the organization of the present work. In Section \ref{secgen} we summarize
the local zeta regularization scheme for the stress-energy VEV of a scalar field
and its connection to the above mentioned kernels, following systematically \cite{FPBook};
in particular, we introduce the \textsl{fundamental operator} $\AAa$ associated to the field
equation and account for the possibility to replace it with the modified version
$\AAae := \AAa + \ee^2$ (depending on the ``infrared cutoff'' $\ee > 0$, which should
be ultimately sent to zero).
In Section \ref{secAA} we consider on $\R^3$ the operator $-\lap$ plus a point delta-type potential
concentrated at the origin; following \cite{AlbB}, we review the rigorous description of this
configuration in terms of the fundamental operator $\AA = -\lap$ on $\Om := \R^3 \setminus \{\b0\}$
(with suitable boundary conditions at the origin) and introduce as well its modified version $\AAae$.
In Section \ref{secHeat} we report an explicit expression for the heat kernel of $\AAae$,
following trivially from a result of \cite{AlbHeat} on the same kernel for $\AAa$;
this expression is rephrased in Section \ref{secSph} in terms of a system of spherical
coordinates, to be used on $\Om$ up to the end of the paper.
\parn
In Section \ref{secDirTmn} we determine the zeta-regularized stress-energy VEV $\la 0|\Tiue_{\mu\nu}|0\ra$
for our field theory with point singularity; more precisely, we derive an integral representation
for this VEV using the previously mentioned expression for the heat kernel and some known relations
involving the Dirichlet kernel of $\AAae$.
This representation of the stress-energy VEV, depending on the regulating parameter $\u \in \C$
and on the infrared cutoff $\ee >0$, is reformulated in Section \ref{secAC} in terms of Bessel
functions; this also allows to determine the analytic continuation of the map
$\u \mapsto \la 0|\Tiue_{\mu\nu}|0\ra$ to a meromorphic function on the whole complex plane,
possessing a simple pole at $\u = 0$.
We compute the regular part of $\la 0|\Tiue_{\mu\nu}|0\ra$ at this point in Section \ref{secRUV}
and subsequently evaluate the limit $\ee \to 0^+$ of the resulting expression in Section \ref{secIR};
according to a general prescription of \cite{FPBook}, these operations determine the renormalized
stress-energy VEV $\la 0|\Ti_{\mu\nu}|0\ra_{ren}$.
The final expressions thus obtained for the non-vanishing components of $\la 0|\Ti_{\mu\nu}|0\ra_{ren}$
are reported in the conclusive Section \ref{secTmn}; therein, we also analyze the asymptotic
behavior of the renormalized stress-energy VEV in various regimes, discussing especially
the expansions for small and large distances from the point impurity
(see, respectively, subsections \ref{subsmallr} and \ref{sublarger}).
\parn
Some of the computations required by this paper were assisted by the symbolic mode of \verb"Mathematica".

\section{The general setting}\label{secgen}
\noindent
\emph{Quantum field theory and the fundamental operator.}
In the present section we briefly recall the general setting of \cite{FPBook}
for the quantum theory of a scalar field on a space domain with boundary conditions,
possibly in presence of a static external potential; this formulation will be
methodically employed in the sequel. \parn
We use natural units, so that $c=1$ and $\hbar=1$, and work in $(1+3)$-dimensional
Minkowski spacetime; this is identified with $\R^4 = \R \times \R^3$ using a set of
inertial coordinates

\begin{equation}
x \,=\, (x^\mu)_{\mu=0,1,2,3} \,\equiv\, (x^0,(x^i)_{i = 1,2,3}) \,\equiv\, (t,\x) ~, \label{xco}
\end{equation}

\noindent
so that the spacetime line element reads

\begin{equation}
d s^2 =\, \eta_{\mu \nu}\, d x^\mu d x^\nu ~, \qquad (\eta_{\mu \nu})\, :=\, \mbox{diag}(-1,1,1,1)~.
\end{equation}

\noindent
We assume that, in this coordinate system,
the spatial domain for the field consists of a fixed open subset $\Om$ of $\R^3$. \parn
To proceed, we consider a canonically quantized, neutral scalar field
$\Fi : \R \times \Om \to \LL_{s a}(\Fock)$, $(t,\x) \mapsto \Fi(t,\x)$
($\Fock$ is the Fock space and $\LL_{s a}(\Fock)$ are the self-adjoint operators on it);
this can interact with a static background potential $V : \Om \to \R$, $\x \mapsto V(\x)$.
We indicate with $|0 \ra \in \Fock$ the vacuum state
and we systematically use the acronym VEV for ``vacuum expectation value''.
We assume the field $\Fi$ to fulfill the Klein-Gordon-like equation

\begin{equation}
(-\,\de_{tt} - \lap + V)\,\Fi \,=\, 0 ~,
\end{equation}

\noindent
with given boundary conditions on $\de\Om$ (here and in the sequel,
$\lap$ is the 3-dimensional Laplacian). The operator

\begin{equation}
\AA \,:=\, -\lap + V
\end{equation}

\noindent
with the prescribed boundary conditions will be called the
\textsl{fundamental operator} of the system. We require $\AA$ to be a self-adjoint,
non-negative operator in $L^2(\Om)$; obviously enough, ``$\AA$ non-negative'' means
that $\AA$ has spectrum $\si(\AA) \subset [0,+\infty)$. These conditions of
self-adjointness and non-negativity are in fact limitations about the admissible
boundary conditions and potentials. \parn
The operator $\AA$ considered in this work corresponds, morally, to a delta-type
potential placed at the origin $\x = \b0$, multiplied by an infinitesimally small
coupling constant. According to the already cited paper of Berezin and Faddeev \cite{BeFa},
this configuration can be described rigorously in terms of the space domain
$\Om := \R^3 \setminus \{\b0 \}$, defining $\AA$ to be the operator $-\lap$ on $\Om$
with suitable boundary conditions at the origin (and with no external potential $V$);
the basic features of $\AA$ will be reviewed in Section \ref{secAA}.
In the remainder of the present Section \ref{secgen} we will not focus on this
specific configuration, referring again to a general field theory as in \cite{FPBook}.
\vspace{0.2cm}

\noindent
\emph{Zeta regularization and renormalization of the stress-energy VEV.}
A quantum field theory of the type considered in \cite{FPBook} is typically affected
by ultraviolet divergences: these appear in the computation of VEVs for many significant
observables, in particular for the stress-energy tensor. To treat these divergences,
one can first regularize the field operator, and then set up a suitable renormalization
procedure; the zeta approach employed in \cite{FPBook} and in the present work is a technique
allowing to achieve these goals. \parn
The field regularization illustrated in \cite{FPBook} requires a self-adjoint, strictly
positive operator on $L^2(\Om)$; the last condition means that the spectrum of the operator
must be contained in $[\ee^2, +\infty)$ for some $\ee >0$. When the fundamental operator
$\AA$ is strictly positive, it can be used directly for the purpose of regularization;
however, in many interesting cases (including the one considered in the present work),
the spectrum $\si(\AA)$ contains a right neighborhood of the zero. In these cases, one
can replace $\AA$ with the modified fundamental operator

\begin{equation}
\AA_{\ee} \,:=\, \AA + \ee^2 \qquad (\ee > 0) \label{aep}
\end{equation}

\noindent
and ultimately take the limit $\ee \to 0^{+}$. The parameter $\ee$ introduced in Eq. \eqref{aep}
can be interpreted as an infrared cutoff; note that $\ee$ is dimensionally a mass in our
units with $c = \hbar = 1$. \parn
After defining the operator \eqref{aep}, we introduce the \textsl{zeta-regularized field operator}

\begin{equation}
\Fiue \,:=\, (\mm^{-2} \AA_{\ee})^{-\u/4}\, \Fi ~, \label{Fiu}
\end{equation}

\noindent
where $\u \in \C$ is the regulating parameter and $\mm > 0$ is a ``mass scale'' parameter;
note that $\Fiue\big|_{\u = 0} = \Fi$, at least formally. We use the above regularized field operator
to define the \textsl{zeta-regularized stress-energy tensor}

\begin{equation}
\Tiue_{\mu \nu} \,:=\, \big(1 - 2\xi\big)\, \de_\mu \Fiue \circ\de_\nu \Fiue
- \l({1\over 2} - 2\xi \r) \eta_{\mu\nu}
\l( \de^\lam\Fiue\,\de_\lam \Fiue + V\,(\Fiue)^2\r)
- 2 \xi \, \Fiue \circ \de_{\mu \nu} \Fiue ~. \label{Tiu}
\end{equation}

\noindent
Here: $\xi \in \R$ is an assigned dimensionless parameter; $A \circ B := (1/2) (A B + B A)$
for all linear operators $A, B$ on $\Fock$; all the bilinear terms in $\Fiue$ are evaluated on
the diagonal (e.g., $\de_\mu \Fiue \circ \de_\nu \Fiue$ indicates the map
$x \in \R \times \Om \mapsto \de_\mu \Fiue(x) \circ \de_\nu \Fiue(x)\,$);
$V\,(\Fiue)^2$ stands for the map $x \equiv (t,\x) \mapsto V(\x)\,\Fiue(x)^2$. \parn
The VEV $\la 0|\Tiue_{\mu\nu}|0\ra$ is well defined and analytic for $\ee > 0$ and $\Re\u$ large enough;
when the map $\u \mapsto \la 0|\Tiue_{\mu\nu}|0\ra$ can be analytically continued to a
neighborhood of $\u = 0$ (possibly, with a singularity at $0$), we define the
\textsl{renormalized stress-energy VEV} as \cite{FPBook}

\begin{equation}
\la 0|\Ti_{\mu \nu}|0\ra_{ren} \,:=\, \lim_{\ee \to 0^{+}} RP\Big|_{\u = 0}\,\la 0|\Tiue_{\mu\nu}|0\ra~,
\label{Tren}
\end{equation}

\noindent
where $RP$ indicates the regular part of the Laurent expansion near $\u = 0$
({\footnote{Consider a complex-valued analytic function $\u \mapsto \FF(\u)$, defined in a
complex neighborhood of $\u = 0$ except, possibly, the origin; then, $\FF$ has Laurent
expansion $\FF(\u) = \sum_{k = -\infty}^{+\infty} \FF_k\,\u^k$.
We define the regular part of $\FF$ near $\u = 0$ to be

$$ (RP\,\FF)(\u) \,:=\, \sum_{k = 0}^{+\infty} \FF_k\,\u^k ~; $$

\noindent
in particular, $RP\big|_{\u = 0} \FF \,=\, \FF_0$~. }}).
Taking the regular part in $\u$ amounts to renormalize the ultraviolet divergences, which are
the harder problem to be solved; then, the cutoff $\ee$ associated to the milder, infrared pathologies
is simply removed taking its zero limit. \parn
For a discussion on the role of the parameter $\xi$ appearing in Eq. \eqref{Tiu}
and in the related VEVs we refer to \cite{FPBook}
(see, especially, Appendix A and references therein). Here we limit ourselves to mention
that the conformal invariance properties of the stress-energy tensor can be discussed and
yield a natural decomposition of the form

\begin{equation}
\la 0|\Ti_{\mu\nu}|0\ra_{ren} \,=\, \Tc_{\mu\nu} + \big(\xi - \xi_c\big)\, \Tnc_{\mu\nu} ~,
\qquad \xi_c \,:=\, {1 \over 6} ~. \label{Tcnc0}
\end{equation}

\noindent
The functions $\Tc_{\mu\nu}$, $\Tnc_{\mu\nu}$ in Eq. \eqref{Tcnc0} are referred to,
respectively, as the conformal and non-conformal parts of the stress-energy VEV
(and $\xi_c$ is called the \textsl{critical value})
({\footnote{
Of course, if we have $\la 0|\Ti_{\mu\nu}|0\ra_{ren}$ for any value of $\xi$,
we obtain its conformal and non-conformal parts with the prescriptions

$$ \Tc_{\mu\nu} \,=\, \la 0|\Ti_{\mu\nu}|0\ra_{ren}\Big|_{\xi = \xi_c} ~, \qquad
\Tnc_{\mu\nu} \,=\, {1 \over \xi_c}\;\Big(\,\Tc_{\mu\nu} - \la 0|\Ti_{\mu\nu}|0\ra_{ren}\Big|_{\xi = 0}\,\Big) ~. $$

}}).
\vspace{0.2cm}

\noindent
\emph{Integral kernels.}
For the implementation of the previous scheme, it is essential to point out the relations
between the regularized stress-energy VEV and some integral kernels \cite{FPBook};
in order to illustrate them, it is convenient to recall some basic facts about such kernels. \parn
In general, given a linear operator $\BB: f \mapsto \BB f$ acting on $L^2(\Om)$,
the \textsl{integral kernel} of $\BB$ is the unique (generalized) function $\Om \times \Om \to \C$,
$(\x,\y) \mapsto \BB(\x,\y)$ such that $(\BB f)(\x) = \int_{\Om} d \y\;\BB(\x,\y)\, f(\y)$\; ($\x \in \Om$). \parn
In particular, let $\Asc$ be a strictly positive self-adjoint operator in $L^2(\Om)$ and
consider the complex power $\Asc^{-s}$, with exponent $s \in \C$; the corresponding kernel
$(\x,\y) \mapsto \Asc^{-s}(\x,\y)$ is called the \textsl{$s$-th Dirichlet kernel} of $\Asc$.
For $\Asc$ strictly positive (or even non negative), we can define the corresponding
heat semigroup $(e^{-\t \Asc})_{\t\,\in\,[0,+\infty)}$; the mapping $(\t,\x,\y) \mapsto e^{-\t \Asc}(\x,\y)$
is called the \textsl{heat kernel} of $\Asc$
({\footnote{The variable $\t$ must not be confused with the time coordinate $t$.}}).
The Mellin-type integral representation
$\Asc^{-s}(\x,\y) =$ $\Ga(s)^{-1}\! \int_0^{+\infty}\! d\t\,\t^{s-1} e^{-\t\Asc}(\x,\y)$
holds true for all $s \in \C$ such that the previous integral converges. \parn
Now, let us return to the quantum field theory of the previous paragraphs;
this has important connections with the Dirichlet and heat kernels of the operator
$\Asc = \AAae$. In fact, it can be shown that the components of the regularized
stress-energy VEV are completely determined by the Dirichlet kernel $\AAae^{-s}(\x,\y)$
via the following relations, where $i,j,\ell \in \{1,2,3\}$ are spatial indexes and summation
over repeated indexes is understood:

\begin{equation}\begin{array}{c}
\dd{\la 0 | \Tiue_{0 0}(\x) | 0 \ra \,=} \vspace{0.15cm}\\
\dd{\mm^\u \l[ \l({1 \over 4} + \xi \r) \AAae^{-{\u - 1\over 2}}(\x,\y)
+ \l({1 \over 4} - \xi \r) \Big(\de^{x^\ell} \de_{y^\ell} + V(\x)\Big)\,
\AAae^{-{\u + 1 \over 2}}(\x,\y)\r]_{\y = \x} \,;}  \label{Tidir00c} \vspace{-0.05cm}
\end{array}\end{equation}

\begin{equation}
\la 0 | \Tiue_{i 0}(\x) | 0 \ra \,=\, \la 0 |\Tiue_{0 i}(\x) | 0 \ra \,=\, 0 ~; \vspace{0.15cm}
\label{Tidiri0c}
\end{equation}

\begin{equation}\begin{array}{c}
\dd{\la 0 | \Tiue_{i j}(\x) | 0 \ra \,=\, \la 0 |\Tiue_{j i}(\x) | 0 \ra \,=} \vspace{0.1cm} \\
\dd{\hspace{-1cm}\mm^\u \l[\l({1\over 4} - \xi\r) \eta_{i j} \l(\AAae^{-{\u - 1 \over 2}}(\x,\y)
- \Big(\de^{\,x^\ell}\de_{y^\ell} + V(\x)\Big)\,\AAae^{-{\u + 1 \over 2}}(\x,\y) \r) + \r.} \vspace{0.1cm} \\
\dd{\hspace{6cm} \l. + \l(\l({1\over 2} - \xi\r)\de_{x^i y^j} - \xi\,\de_{\!x^i x^j}\r)
\AAae^{-{\u + 1 \over 2}}(\x,\y)\, \r]_{\y = \x}}  \label{Tidirijc}
\end{array}\end{equation}

\noindent
($\la 0|\Tiue_{\mu \nu}(\x)|0\ra$ is short for $\la 0|\Tiue_{\mu\nu}(t,\x)|0\ra$;
indeed, the VEV does not depend on the time coordinate $t$). In a number of interesting cases,
explicit expressions are available for the heat kernel $e^{-\t \AA}(\x,\y)$ of the fundamental
operator $\AA$. Recalling Eq. \eqref{aep}, we obtain from here the heat kernel of the modified
fundamental operator $\AAae$ via the identity

\begin{equation}
e^{-\t \AAae}(\x,\y) \,=\, e^{-\ee^2 \t}\; e^{-\t \AA}(\x,\y) ~. \label{heregen}
\end{equation}

\noindent
Subsequently, we can determine the Dirichlet kernels appearing in
Eq.s \eqref{Tidir00c}-\eqref{Tidirijc} via the Mellin relation

\begin{equation}
\AAae^{-s}(\x,\y) \,=\, {1 \over \Ga(s)} \int_0^{+\infty}\!\!
d\t\;\t^{s-1}\,e^{-\t\AAae}(\x,\y) ~. \vspace{0.2cm} \label{Melgen}
\end{equation}

\noindent
\emph{Curvilinear coordinates.}
In order to fit the symmetries of the specific problem under analysis, it is often useful
to consider on $\Om$ a set of curvilinear coordinates $\q \equiv (q^i)_{i = 1,2,3}$ in place of the
Cartesian coordinates $\x = (x^i)$; this induces a set of coordinates $q \equiv (q^\mu) \equiv (t,\q)$
on Minkowski spacetime. The line elements of $\Om$ and of Minkowski spacetime read, respectively,

\begin{equation}
d \ell^2 \,=\, a_{i j}(\q)\,dq^i dq^j ~, \qquad
d s^2 \,=\, - dt^2 + d \ell^2 \,=\, g_{\mu \nu}(q)\, d q^\mu d q^\nu ~,
\end{equation}

\noindent
where $a_{i j}(\q)$ is a suitable symmetric and positive definite matrix, while

\begin{equation}
g_{0 0}(q) \,:=\, -1 ~, \qquad g_{0 i}(q) \,=\, g_{i 0}(q) \,:=\, 0 ~,
\qquad g_{i j}(q) \,:=\, a_{i j}(\q) ~.
\end{equation}

\noindent
For the components of the stress-energy tensor in the spacetime coordinates $q^\mu$
we have an expression similar to \eqref{Tiu}, with $\eta_{\mu\nu}$ and the second order
derivatives $\de_{\mu \nu}$ replaced, respectively, by the metric coefficients
$g_{\mu \nu}(\q)$ and by the corresponding covariant derivatives $\nabla_{\!\mu \nu}$
({\footnote{Let us recall that the first order covariant derivatives $\nabla_{\!\mu}$
coincide with the ordinary derivatives $\de_{\mu}$ on scalar functions.}}). \parn
Obviously enough, a function $\x \mapsto f(\x)$ on $\Om$ or $(\x,\y) \mapsto h(\x,\y)$ on
$\Om \times \Om$ induces a function of the curvilinear coordinates $\q$ of $\x$
and $\p$ of $\y$; we indicate the latter function with the slightly abusive notation
$\q \mapsto f(\q)$ or $(\q,\p) \mapsto h(\q,\p)$. Keeping this in mind,
we can write the following analogues of Eq.s \eqref{Tidir00c}-\eqref{Tidirijc}
\cite{FPBook}:

\begin{equation}\begin{array}{c}
\dd{\la 0 | \Tiue_{0 0}(\q) | 0 \ra \,=} \vspace{0.15cm}\\
\dd{ \mm^\u \l[ \l({1 \over 4} + \xi \r) \AAae^{-{\u - 1\over 2}}(\q,\p)
+ \l({1 \over 4} - \xi \r) \Big(\de^{q^\ell} \de_{p^\ell} + V(\q)\Big)\,
\AAae^{-{\u + 1 \over 2}}(\q,\p)\r]_{\p = \q} \,;} \label{Tidir00} \vspace{-0.05cm}
\end{array}\end{equation}

\begin{equation}
\la 0 | \Tiue_{i 0}(\q) | 0 \ra \,=\, \la 0 |\Tiue_{0 i}(\q) | 0 \ra \,=\, 0 ~;
\vspace{0.15cm}\label{Tidiri0}
\end{equation}

\begin{equation}\begin{array}{c}
\dd{\la 0 | \Tiue_{i j}(\q) | 0 \ra \,=\, \la 0 |\Tiue_{j i}(\q) | 0 \ra \,=} \vspace{0.15cm}\\
\dd{\hspace{-1cm} \mm^\u \l[\l({1\over 4} - \xi\r) a_{i j}(\q) \l(\AAae^{-{\u - 1 \over 2}}(\q,\p)
- \Big(\de^{\,q^\ell}\de_{p^\ell} + V(\q)\Big)\, \AAae^{-{\u + 1 \over 2}}(\q,\p)\r) + \r.}\vspace{0.1cm}\\
\dd{\hspace{6cm} \l. + \l(\l({1\over 2} - \xi\r)\de_{q^i p^j} - \xi\,D_{q^i q^j} \r)
\AAae^{-{\u + 1 \over 2}}(\q,\p)\, \r]_{\p = \q}} ~. \label{Tidirij}
\end{array}\end{equation}

\noindent
In the above, $D_{q^i q^j}$ are the covariant derivatives of second order corresponding
to the metric coefficients $a_{i j}(\q)$ of the given curvilinear coordinates on $\Om$;
let us recall that, for any scalar function $f$ on $\Om$, we have

\begin{equation}
D_{q^i q^j}f \,=\, \de_{q^i q^j} f - \ga^{k}_{ij}(\q)\; \de_{q^k} f \label{compu}
\end{equation}

\noindent
where $\ga^{k}_{ij}$ are the Christoffel symbols for the spatial metric $a_{i j}$, i.e.,
$\ga^{k}_{ij} = {1 \over 2}\, a^{k\ell} (\de_{i}a_{\ell j} + \de_{j} a_{i \ell} - \de_{\ell}a_{ij})$.
Of course, the analogues of Eq.s \eqref{heregen} \eqref{Melgen} in curvilinear coordinates are

\begin{equation}
e^{-\t \AAae}(\q,\p) \,=\, e^{-\ee^2 \t}\; e^{-\t \AA}(\q,\p)~, \label{here}
\end{equation}

\begin{equation}
\AAae^{-s}(\q,\p) \,=\, {1 \over \Ga(s)} \int_0^{+\infty}\!\!
d\t\;\t^{s-1}\,e^{-\t\AAae}(\q,\p) ~. \label{Mel}
\end{equation}

\section{The fundamental operator for a point impurity}\label{secAA}
The precise definition of the operator $\AA$ corresponding to a delta-type potential
is a non-trivial problem, whose treatment depends crucially on the co-dimension
of the support of the delta-type potential.
As already indicated, the case of a point impurity in spatial dimension $d = 3$ (with support of co-dimension $3$)
was first treated in a mathematically precise setting by Berezin and Faddeev in \cite{BeFa}.
These authors proposed an approach to define the operator

\begin{equation}
``~ \AA \,:=\, - \lap + \l(\be + {\be^2 \over 4 \pi \at}\r) \del_\b0 ~ " ~, \label{oper}
\end{equation}

\noindent
where $\del_{\b0}$ is the Dirac delta at the origin, $\at \in \R \setminus \{0 \}$ is
a fixed parameter and $\be$ is infinitesimally small; we already mentioned that
the infinitesimal nature of the coupling constant can be interpreted as the effect of a renormalization.
The approach of \cite{BeFa} was refined in many subsequent works. Here we mention,
in particular, the book \cite{AlbB} by Albeverio \textsl{et al.}
({\footnote{The present variable $\at$ is connected to the variable $\alpha$ of \cite{AlbB}
by the relation $\at = 1/(4 \pi \alpha)$\,.}}); see also the vast literature cited therein. \parn
According to the references mentioned above, the heuristic expression \eqref{oper} has a rigorous
counterpart based on the space domain

\begin{equation}
\Om \,:=\, \R^3 \setminus \{\b0 \} \label{Ompap}
\end{equation}

\noindent
and on the Laplacian on this domain, with an appropriate boundary condition at the origin. \parn
To define precisely this counterpart, from now on we intend the derivatives, the Laplacian,
etc.\! of functions on $\R^3$ (or on $\Om$) in the sense of the Schwartz distribution theory.
We indicate with $H^2(\R^3)$ the Sobolev space of complex-valued functions on $\R^3$
whose (distributional) derivatives up to second order are in $L^2(\R^3)$; we recall that
$H^2(\R^3)$ is embedded in the space $C_B(\R^3)$ of bounded, continuous functions on $\R^3$
\cite{Adams}. \parn
To go on, for each $z \in \C \setminus [0,+\infty)$ we consider the function
({\footnote{Here and in the following, we consider the principal determination of the argument
for complex numbers, i.e., $\arg : \C \setminus [0,+\infty) \to (0,2\pi)$; furthermore,
for any $z \in \C \setminus [0,+\infty)$, we always write $\sqrt{z}$ to indicate the square root
determined by this choice of the argument, i.e., the one with $\Im \sqrt{z} > 0$\,.}})

\begin{equation}
\G_z \,:\, \Om \to \C~, \qquad \G_z(\x) \,:=\, {e^{i \sqrt{z}\;|\x|} \over 4\pi\,|\x|}~;
\end{equation}

\noindent
note that $\G_z \in L^2(\Om)$ and that $(-\lap -z)\,\G_z = 0$ everywhere in $\Om$
({\footnote{Note that, even though $(-\lap -z)\,\G_z = 0$ in $\Om$,
one has $(-\lap -z)\,\G_z = \del_{\b0}$ in $\R^3$; this shows, in particular, that
$\G_z$ does not belong to $H^2(\R^3)$.}}).
Then, after fixing $\at \in \R$ we set

\begin{equation}\begin{array}{c}
\dd{\Dom\AAa \,:=\, \l\{\psi \in L^2(\Om) \,\l|~\exists\,
z \!\in\! \C \setminus [0,+\infty),\, \vfi \!\in\! H^2(\R^3) ~~\mbox{s.t.}~~
\psi = \vfi + {4 \pi \at \over 1 -  i\sqrt{z}\,\at}\; \vfi(\b0)\; \G_z \r.\r\} ,} \vspace{0.2cm} \\
\dd{\AAa \,:=\, (-\lap) \rist \Dom\AAa \subset L^2(\Om) \to L^2(\Om) ~. \label{defAAa}}
\end{array}\end{equation}

\noindent
Let us point out some known facts about the operator $\AA$ defined above.
\vspace{0.05cm}\parn
\noindent
i) The condition characterizing a function $\psi$ in the domain of $\AA$ is in fact
a boundary condition at the origin $\x = \b0$: $\psi$ is required to be the sum of
a function $\vfi \in H^2(\R^3) \subset C_B(\R^3)$, well defined even at the origin,
and of another function diverging at the origin, with the peculiar form
${4 \pi \at \over 1 -  i\sqrt{z}\,\at}\; \vfi(\b0)\; \G_z$\,. In addition, for any
fixed $z \in \C \setminus [0,+\infty)$, this decomposition of $\psi$ is shown to be
unique \cite{AlbB,PosKre}.
\vspace{0.05cm} \parn
\noindent
ii) Consider a function $\psi \in \Dom \AA$ and its decomposition as in \eqref{defAAa},
based on some pair $(z,\vfi)$. For any $z'\in \C \setminus [0,+\infty)$,
$\psi$ has a similar representation based on the pair $(z',\vfi')$, where
$\vfi' = \vfi + {4 \pi \at \over 1 - i \sqrt{z}\,\at}\;\vfi(\b0)\,(\G_z-\G_{z'})$
({\footnote{Let us remark that the difference $\G_z - \G_{z'}$ does indeed belong
to the Sobolev space $H^2(\R^3)$, despite the fact that $\G_z$ and $\G_{z'}$ are both singular at the origin.
To prove this claim it suffices to recall that $\G_{z'} \in L^2(\Om) \simeq L^2(\R^3)$ and to use
the resolvent-type identity $\G_z - \G_{z'} = (z'\!-z)\, R_0(z)\,\G_{z'}$ (see, e.g., Lemma 2.1 of \cite{PosKre}),
where the bounded operator $R_0(z) : L^2(\R^3) \to H^2(\R^3)$ is the resolvent associated to
the free Laplacian $(-\lap) \rist H^2(\R^3)$.}})\,.
\vspace{0.05cm} \parn
\noindent
iii) Consider again a decomposition as in \eqref{defAAa} for a function $\psi \in \Dom \AA$;
recalling that \mbox{$(-\lap -z)\,\G_z = 0$}, we have $(-\lap -z)\,\psi = (-\lap - z)\,\vfi$ in $\Om$.
This identity if often used in manipulations involving $\AA$; incidentally, the expression on
the right-hand side is in $L^2(\Om)$ (since $\vfi \in H^2(\R^3)$), which ensures $(-\lap - z)\,\psi$
and $-\lap \psi = \AA \psi$ to be as well in $L^2(\Om)$, as stated in the definition \eqref{defAAa}.
\vspace{0.2cm} \parn
The analysis performed in \cite{AlbB,BeFa} shows that the setting on $\Om$ based on the operator
\eqref{defAAa} is morally equivalent (for $\at \neq 0$) to the configuration suggested by Eq. \eqref{oper}.
Let us remark that the prescription \eqref{defAAa} with $\at=0$ gives

\begin{equation}
\Dom \AA \,\big|_{\at = 0} \,=\, H^2(\R^3)~; \label{defAz}
\end{equation}

\noindent
this shows, in particular, that the fundamental operator $\AA$ coincides with the free Laplacian
\mbox{$(-\lap)\!\rist\!H^2(\R^3)$} for $\at = 0$. \parn
Concerning the spectrum of $\AAa$, we refer to Theorem 1.1.4 of \cite{AlbB}.
For each $\at \in \R$, the continuous spectrum of $\AAa$ is in fact absolutely continuous and

\begin{equation}
\si_c(\AAa) \,=\, \si_{ac}(\AAa) \,=\, [0,+\infty) ~; \label{sicAAa}
\end{equation}

\noindent
in this regard, let us mention that the scattering theory for $\AAa$ developed in Section I.1.4
of the same book allows to interpret $-\at$ as the $s = 0$, partial wave scattering length.
Referring to the point spectrum of $\AAa$, we have

\begin{equation}
\si_p(\AAa) \,=\, \l\{\! \begin{array}{ll}
\dd{\emptyset}          & ~~\dd{\mbox{if $\at \geqs 0$}~,} \vspace{0.1cm}\\
\dd{\{ - 1/\at^2 \}}    & ~~\dd{\mbox{if $\at < 0$}~.}
\end{array}\r. \label{sipAAa}
\end{equation}

\noindent
The appearance of a negative eigenvalue for $\at <0$
prevents the perturbed operator $\AAa$ from fulfilling the basic assumption of non-negativity,
which is necessary in order to set up a field theory in the framework of \cite{FPBook};
for this reason, throughout this work we restrict the attention to the sole case

\begin{equation}
\at \geqs 0 ~,
\end{equation}

\noindent
where $\si(\AA) = [0,+\infty)$. \parn
\textsl{From here to the end of the paper, $\Om$ is the space domain \eqref{Ompap} and $\AA$
is the operator \eqref{defAAa} for some fixed $\at \geqs 0$}. We consider a field theory on $\Om$,
with fundamental operator $\AA$; since $\AA$ is just the (opposite of the) Laplacian on this domain,
we will apply the setting of Section \ref{secgen} with $V(\x) = 0$ for all $\x \in \Om$.
Of course, the equivalent of this statement in any curvilinear coordinate system $\q$ for $\Om$ is

\begin{equation}
V(\q) \,=\, 0~. \label{withthe}
\end{equation}

\noindent
Since $\si(\AA)$ contains a right neighborhood of zero, following Eq. \eqref{aep}, we will introduce
an infrared cutoff $\ee > 0$ and consider the modified fundamental operator $\AAae := \AA + \ee^2$
in place of $\AA$; at the end of the paper (see, in particular, Section \ref{secIR}), $\ee$ will
be sent to zero.

\section{The heat kernel for a point impurity}\label{secHeat}
The heat kernel of $\AA$ has been computed in \cite{AlbHeat} (see, in particular, Eq. (3.4) on page 228);
from this result and from Eq. \eqref{here} we obtain the following, for $\x,\y \in \Om$:

\begin{equation}\begin{array}{c}
\dd{e^{-\t\AAae}(\x,\y) = } \vspace{0.15cm}\\
\dd{{e^{-\,\ee^2 \t} \over (4\pi \t)^{3/2}} \!\l[
e^{-{|\x-\y|^2 \over 4\t}}
+ {2\,\t \over |\x|\,|\y|}\l(e^{-{(|\x|+|\y|)^2 \over 4\t}}
- {1 \over \at} \int_0^{+\infty}\!\! d\ww\;e^{-\l({\ww \over \at}+{(\ww + |\x|+|\y|)^2 \over 4\t}\r)}\r)\r].}
\end{array} \label{heatexp}
\end{equation}

\noindent
In passing, let us notice that the above expression for the heat kernel can be viewed
as the sum of two distinct terms. The first one coincides with the standard
heat kernel associated to the modified, free operator $-\lap + \ee^2$
({\footnote{Indeed, let us recall that in spatial dimension $d = 3$ the heat kernel
associated to the operator $-\lap+\ee^2$ on $H^2(\R^3)$ has the form ($\x,\y \in \R^3$)

$$ e^{-\t(-\lap + \ee^2)}(\x,\y) \,=\, {e^{-\,\ee^2\t} \over (4\pi \t)^{3/2}}\;e^{-{|\x-\y|^2 \over 4\t}} ~. $$

}});
for this reason, it can be viewed as a ``free-theory'' contribution, which also appears
when $\at = 0$. The second term corresponds to the two addenda within the
round brackets in Eq. \eqref{heatexp}; this term can be viewed as a ``perturbative''
contribution and it can be easily checked that it vanishes for $\at \to 0^+$.

\section{Spherical coordinates}\label{secSph}
To fit the symmetries of the problem under analysis, let us consider on $\Om$ the standard
spherical coordinates $\q = (r,\te,\vfi) \in (0,+\infty) \times (0,\pi) \times (0,2\pi)$,
which are related to the Cartesian coordinates $\x$ by

\begin{equation}
x^1 =\, r\,\sin\te\,\cos\vfi ~, \qquad
x^2 =\, r\,\sin\te\,\sin\vfi ~, \qquad
x^3 =\, r\,\cos\te ~.
\end{equation}

\noindent
Of course, the metric coefficients in spherical coordinates are
$(a_{i j}(\q)) = \mbox{diag}(1, r^2, r^2 \sin^2 \te)$, and the
corresponding Christoffel symbols are readily obtained. Now, let

\begin{equation}
\q = (r,\te,\vfi) ~, \qquad \p = (r',\te',\vfi')~; \label{bqbp}
\end{equation}

\noindent
then, the correspondent of Eq. \eqref{heatexp} in spherical coordinates reads

\begin{equation}\begin{array}{c}
\dd{e^{-\t\AAae}(\q,\p) \,= } \vspace{0.15cm} \\
\dd{{e^{-\t\,\ee^2} \over (4\pi \t)^{3/2}} \!\l[
e^{-{r^2+{r'}^2 -\,2\,r\,r'S(\te,\te',\vfi-\vfi') \over 4\t}}
+\, {2\,\t \over r\,r'}\l(e^{-{(r+r')^2 \over 4\t}}
- {1 \over \at} \int_0^{+\infty}\!\! d\ww\;
e^{-\l({\ww \over \at}+{(\ww + r+r')^2 \over 4\t}\r)}\r)\r],} \label{heatcu1}
\end{array}\end{equation}

\noindent
where

\begin{equation}
S(\te,\te',\vfi-\vfi') \,:=\, \cos(\te-\te')\,\cos^2\!\l({\vfi-\vfi' \over 2}\r)
+\,\cos(\te+\te')\,\sin^2\!\l({\vfi-\vfi' \over 2}\r) ~. \label{heatcu2}
\end{equation}

\noindent
note that $r^2\!+{r'}^2\! - 2\,r\,r'S(\te,\te',\vfi-\vfi')$ is just
the expression of $|\x - \y|^2 = |\x|^2\! + |\y|^2\! - 2\,\x \cdot \y$ when
$\x,\y$ have spherical coordinates $\q,\p$ as in Eq. \eqref{bqbp}.

\section{The regularized stress-energy VEV}\label{secDirTmn}
Let us keep the coordinate system and the notations of the previous section.
We recall that, for (suitable) $s \in \C$, the $s$-th Dirichlet kernel $\AAae^{-s}(\q,\p)$
can be expressed via Eq. \eqref{Mel}; the integral over $\t$ appearing therein
involves the heat kernel $e^{-\t\AAae}(\q,\p)$ given by Eq.s \eqref{heatcu1} \eqref{heatcu2},
which, in turn, comprises an integral over another variable $\ww$. In the end,
we obtain an explicit representation for $\AAae^{-s}(\q,\p)$, containing integrals
for $\t, \ww \in (0,+\infty)$. \parn
It is readily inferred that the above mentioned integral representation of $\AAae^{-s}(\q,\p)$
is well defined, even along the diagonal $\p = \q$, for any $s \in \C$ with $\Re s > 3/2$
({\footnote{Notice that, as usual, the restriction on $\Re s$ descends from the behavior
of the integrand function for $\t \to 0^+$. On the other hand, let us remark that the presence
of the infrared cutoff parameter $\ee > 0$ is essential in order to ensure the convergence
of the integral for large values of $\t$ (for any $s \in \C$).}}).
By differentiation, we obtain analogous representations for the first order derivatives in $\q,\p$
and for the second order covariant derivatives in $\q$ of the Dirichlet kernel; on the diagonal
$\p = \q$, these representations always make sense for $\Re s$ sufficiently large. \parn
To proceed, let us consider the relations \eqref{Tidir00}-\eqref{Tidirij} (and \eqref{withthe}),
allowing to express the
VEV of the zeta-regularized stress-energy tensor in terms of the Dirichlet kernel $\AAae^{-s}(\q,\p)$.
Using the integral representations discussed formerly for the Dirichlet kernel $\AAae^{-s}(\q,\p)$
and for its derivatives, we obtain the forthcoming explicit expressions \eqref{T00Int}-\eqref{T33Int}
for the non-vanishing components of the zeta-regularized stress-energy VEV. These expressions
are derived introducing, for any fixed $r > 0$, the new integration variables
$\vv := \ww/(2\,r) \in (0,+\infty)$ and $\tt := \t/r^2 \in (0,+\infty)$:

\begin{equation}\begin{array}{c}
\dd{\la 0 | \Tiue_{0 0}(\q) | 0 \ra ~= } \vspace{0.15cm}\\
\dd{{\mm^\u \over (4\pi)^{3/2}\, \Ga({\u + 1\over 2})\,r^{4-\u}}
\int_0^{+\infty}\!\! d\tt\;\tt^{{\u \over 2} - 3}\; e^{-\ee^2 r^2 \tt}\,
\Bigg[\l({1 \over 4} - 2\xi \r) + \l({1 \over 4} + \xi \r) {\u \over 2}\; +} \vspace{0.15cm}\\
\dd{ + \l(\l({1 \over 2} - 2\xi \r) \big(\tt^2 + 1\big)
+ \l({1 \over 2} - 4\xi \r)\tt + \l({1 \over 4} + \xi \r)\tt\;\u \r) e^{-{1/\tt}}\;+} \vspace{0.15cm}\\
\dd{-\;{2r \over \at} \int_0^{+\infty}\!\! d\vv\;e^{-\l({1 \over \tt}\,(\vv + 1)^2\,+\,{2r \over \at}\,\vv\r)}
\l(\!\l({1 \over 2} - 2 \xi\r) \big(\tt + \vv + 1\big)^2 - {\tt \over 2}
+\l({1 \over 4} + \xi\r)\tt\,\u \r)\Bigg] ~;} \label{T00Int}
\end{array}\end{equation}

\begin{equation}\begin{array}{c}
\dd{\la 0 | \Tiue_{r r}(\q) | 0 \ra  ~= } \vspace{0.15cm}\\
\dd{{\mm^\u \over (4\pi)^{3/2}\, \Ga({\u + 1\over 2})\,r^{4-\u}}
\int_0^{+\infty}\!\! d\tt\;\tt^{{\u \over 2} - 3}\; e^{-\ee^2 r^2 \tt}\,
\Bigg[\!-\l({1 \over 4} - 2\xi\r) + \l({1 \over 4} - \xi\r){\u \over 2}\; +} \vspace{0.15cm}\\
\dd{ + \l(\l({1 \over 2} - 4 \xi\r)\tt^2 + \l({1 \over 2} - 2 \xi\r)(\tt+1)
+ \l({1 \over 4} - \xi \r)\tt\;\u \r) e^{-{1/\tt}}\;+} \vspace{0.15cm}\\
\dd{\!-\;{2r \over \at}\!\int_0^{+\infty}\!\!\! d\vv\;e^{-\l({1 \over \tt}\,(\vv + 1)^2+\,{2r \over \at}\,\vv\r)}\!
\l(\!\l({1 \over 2} - 2 \xi\r)\! \big((\tt + \vv + 1)^2\! - \tt\big)
- 2 \xi\,\tt^2 + \l({1 \over 4} - \xi\r) \tt\, \u \r)\!\Bigg]\,;\!}
\end{array}\end{equation}

\begin{equation}\begin{array}{c}
\dd{\la 0 | \Tiue_{\te\te}(\q) | 0 \ra  ~= }\vspace{0.15cm}\\
\dd{{\mm^\u \over (4\pi)^{3/2}\, \Ga({\u + 1\over 2})\,r^{2-\u}}
\int_0^{+\infty}\!\! d\tt\;\tt^{{\u \over 2} - 3}\; e^{-\ee^2 r^2 \tt}\,
\Bigg[\!-\l({1 \over 4} - 2\xi\r) + \l({1 \over 4} - \xi\r){\u \over 2}\; +} \vspace{0.15cm}\\
\dd{-\l(\l({1 \over 2} - 4 \xi\r)(\tt+1)\,\tt
+ \l({1 \over 2} - 2 \xi\r)(\tt+1)
- \l({1 \over 4} - \xi \r)\tt\;\u \r) e^{-{1/\tt}}\;+} \vspace{0.15cm}\\
\dd{\!\!+\;{2r \over \at}\!\int_0^{+\infty}\!\!\!\! d\vv\,e^{-\l({1 \over \tt}(\vv + 1)^2+\,{2r \over \at}\,\vv\r)}\!
\l(\!\l({1 \over 2} - 2 \xi\r) (\tt\!+\!\vv\!+\!1)^2
- 2 \xi (\tt\!+\!\vv\!+\!1)\tt - \l({1 \over 4} - \xi\r)\tt\, \u\! \r)\!\Bigg].\!\!\!} \label{T22Int}
\end{array}\end{equation}

\noindent
Moreover, in compliance with the spherical symmetry of the problem under analysis, we have

\begin{equation}
\la 0 | \Tiue_{\vfi\vfi}(\q) | 0 \ra \,=\, \sin^2\!\te\; \la 0 | \Tiue_{\te\te}(\q) | 0 \ra ~. \label{T33Int}
\end{equation}

\noindent
Consistently with the facts mentioned before about the integral representation of the
Dirichlet kernel (and of its derivatives), it can be checked by direct inspection that all
the integrals appearing in Eq.s \eqref{T00Int}-\eqref{T22Int} are finite for any fixed $r,\ee >0$
and for all complex $\u$ with

\begin{equation}
\Re\u > 4 ~; \label{Reu}
\end{equation}

\noindent
moreover, the maps $\u \mapsto \la 0 | \Tiue_{\mu\nu}(\q) | 0 \ra$
($\mu,\nu \in \{0,r,\te,\vfi\}$) described by Eq.s \eqref{T00Int}-\eqref{T33Int}
are analytic in the region \eqref{Reu}. In the following Section \ref{secAC}, we
re-express the previous results in terms of Bessel functions; this automatically
gives the analytic continuations of the maps $\u \mapsto \la 0 | \Tiue_{\mu\nu}(\q)| 0 \ra$,
which are meromorphic functions on the whole complex plane with simple poles.
Such continuations will be used in the subsequent Sections \ref{secRUV}, \ref{secIR} to determine
the renormalized stress-energy VEV; for brevity, we shall give the details of these computations
only for the map $\u \mapsto \la 0 | \Tiue_{00}(\q) | 0 \ra$, which is related to the
energy density.

\section{Expressing the previous results via Bessel functions; analytic continuation}\label{secAC}
Let us consider the representation \eqref{T00Int} for the component $\la 0|\Tiue_{00}(\q)|0\ra$
of the regularized stress-energy VEV, involving integrals over the two variables
$\tt,\vv \in (0,+\infty)$. It can be easily checked that, for any $\u \in \C$ with $\Re u > 4$
(see Eq. \eqref{Reu}), the order of integration over these variables can be interchanged due to
Fubini's theorem.\\
On the other hand, let us point out the following relations, descending from well-known
integral representations for the Euler Gamma function $\Ga$ and for the modified Bessel function
of second kind $K_\si$ (see, respectively, Eq.s 5.9.1 and 10.32.10 of \cite{NIST}):

\begin{equation}
\int_0^{+\infty}\!\!d\tt\;\tt^{\si - 1}\, e^{-a^2\tt} \,=\, a^{-2\si}\; \Ga(\si) \qquad
\mbox{for all $a > 0$, $\si \in \C$ with $\Re \si > 0$} ~; \label{71}
\end{equation}

\begin{equation}
\int_0^{+\infty}\!\!d\tt\;\tt^{\si - 1}\,e^{-a^2\tt - {b^2 \over \tt}} \,=\,
2 \l({b \over a}\r)^{\!\si} K_{-\si}(2\,a\,b) \qquad
\mbox{for all $a,b > 0$, $\si \in \C$} ~. \label{eqK}
\end{equation}

\noindent
In view of the developments to be discussed in the forthcoming Sections \ref{secRUV} and \ref{secIR},
for any $\si \in \C$ it is advantageous to consider in place of the Bessel function $K_\si$ the map

\begin{equation}
\KM_{\si} : (0,+\infty) \to \C\,, \qquad \zz \mapsto \KM_{\si}(\zz) := \zz^{\si}\,K_{\si}(\zz)\;;
\label{KMdef}
\end{equation}

\noindent
using this function, Eq. \eqref{eqK} can be rephrased as

\begin{equation}
\int_0^{+\infty}\!\!d\tt\;\tt^{\si - 1}\,e^{-a^2\tt - {b^2 \over \tt}} \,=\,
2^{\si+1}\, b^{2\si}\,\KM_{-\si}(2\,a\,b) \qquad
\mbox{for all $a,b > 0$, $\si \in \C$} ~. \label{eqKM}
\end{equation}

\noindent
Using Eq.s \eqref{71} \eqref{eqKM}, by a few additional algebraic manipulations
we obtain from Eq. \eqref{T00Int} that \vspace{0.2cm}\parn

\vbox{

\begin{equation}
\la 0 | \Tiue_{0 0}(\q) | 0 \ra ~=~
{\ee^4\; \Ga\!\l({\u \over 2}-2\r) \over (4\pi)^{3/2}\; \Ga({\u + 1\over 2})} \l({\mm \over \ee}\r)^{\!\u}
\l[\l({1 \over 4} - 2\xi \r) + \l({1 \over 4} + \xi \r) {\u \over 2}\,\r] + \label{T00II} \vspace{-0.1cm}
\end{equation}

\begin{equation*}\begin{array}{c}
\dd{ +\, {2^{{\u \over 2}}\,\mm^\u \over (4\pi)^{3/2}\, \Ga({\u + 1\over 2})\,r^{4-\u}}\,
\Bigg[\!\l({1 \over 4} - \xi\r) \KM_{2-{\u \over 2}}(2\,\ee\,r)
+\! \l(\!\l({1 \over 2} - 4\xi\r)\! + \!\l({1 \over 4} + \xi \r) \u\r)
\KM_{1-{\u \over 2}}(2\,\ee\,r)\;+} \vspace{0.15cm}\\
\dd{+ \,\big(1 - 4\xi\big)\, \KM_{-{\u \over 2}}(2\,\ee\, r)
-\,{2r \over \at} \int_0^{+\infty}\!\! d\vv\; {e^{- {2r \over \at}\,\vv} \over (\vv + 1)^{2-\u}}\,
\Bigg(\!\big(1 - 4 \xi\big)\, (\vv + 1)^2\,\KM_{-{\u \over 2}}\big(2\,\ee\,r\,(\vv + 1)\big)\;+ } \\
\dd{+ \l( (1 - 4 \xi)\, (\vv + 1) - {1 \over 2} +\!\l({1 \over 4} + \xi\r) \u \r)
\KM_{1-{\u \over 2}}\big(2\,\ee\,r\,(\vv + 1)\big)
+ \l({1 \over 4} - \xi\r)\KM_{2-{\u \over 2}}\big(2\,\ee\,r\,(\vv + 1)\big)\!\Bigg) \Bigg]\,.}
\end{array}\end{equation*}

}

\noindent
Even though the above identity was derived under the restriction $\Re \u > 4$
on the regulating parameter $\u$, we claim that Eq. \eqref{T00II} automatically determines
the analytic continuation of the map $\u \mapsto \la 0 | \Tiue_{0 0}(\q) | 0 \ra$ to a function
which is meromorphic on the whole complex plane, with only simple poles.
In the following items (i)-(iii) we briefly account for the last statement.\parn
\noindent
i) The reciprocal of the Euler Gamma function $\Ga$ is analytic on the whole complex plane
(see, e.g., \S 5.2(i) of \cite{NIST}); so the Gamma's in the denominators of Eq. \eqref{T00II}
give no problem from the viewpoint of analyticity. \parn
\noindent
ii) From the analyticity properties of the Gamma function (see, again, \S 5.2(i) of \cite{NIST})
it can be readily inferred that the term in the first line of Eq. \eqref{T00II}
is a meromorphic function of $\u$, with simple poles at

\begin{equation}
\u \in \{4,2,0,-2,-4,\,...\,\} ~, \label{poles}
\end{equation}

\noindent
(where the argument of the Gamma function in the numerator of the above mentioned term
is a non-positive integer).
In passing, let us remark that the expression under analysis does not depend on
$r$ or $\at$; indeed, this terms descends solely from the ``free-theory'' contribution to the heat kernel
(see the comments below Eq. \eqref{heatexp}). \parn
\noindent
iii) Let us now consider the terms in the second, third and fourth line of Eq. \eqref{T00II}.
From some basic properties of the modified Bessel function $K_\si$
(see, e.g., \S 10.25(ii), \S 10.38 and \S 10.40 of \cite{NIST}) we infer that the function
$\KM_\si$ defined in Eq. \eqref{KMdef} has the following regularity features: for any fixed
$\zz \in (0,+\infty)$, the map $\si \mapsto \KM_\si(\zz)$ is analytic on the whole complex plane;
for any fixed $\si \in \C$, both the maps $\zz \mapsto \KM_\si(\zz)$ and $\zz \mapsto (\de \KM_\si/\de\si)(\zz)$ are
analytic (whence, in particular, continuous) for $\zz \in (0,+\infty)$ and they decay exponentially
for $\zz \to +\infty$. The facts mentioned above about $\KM_\si$ and $\Ga$
suffice to infer that the terms under analysis determine an analytic function of the regulating
parameter $\u$, defined on the whole complex plane.
\parn
\vspace{0.05cm}
Before proceeding, let us remark that analogous results can be derived for the analytic
continuations of the maps $\u \mapsto \la 0 | \Tiue_{\mu\nu}(\q) | 0 \ra$, associated to the
other components of the regularized stress-energy VEV.
In the forthcoming Sections \ref{secRUV}, \ref{secIR} we determine the renormalized VEV
of the stress-energy tensor, starting from these analytic continuations and implementing
the definition \eqref{Tren} of $\la 0|\Ti_{\mu \nu}(\q)|0\ra_{ren}$.


\section{Renormalization of ultraviolet divergences:
the regular part at \boldmath{$\u = 0$} }\label{secRUV}
The results reported in the previous section show, amongst else, that the analytic continuations
of the maps $\u \mapsto \la 0|\Tiue_{\mu\nu}(\q)|0\ra$ possess a simple pole at the
point $\u = 0$ (see, e.g., Eq. \eqref{poles}), of interest for renormalization.
In the present section we proceed to determine the corresponding regular part at $\u = 0$,
appearing in the definition \eqref{Tren} of the renormalized stress-energy VEV. As an example,
we shall report here the details of the related computations only for the energy density component
$RP\big|_{\u = 0}\la 0|\Tiue_{00}(\q)|0\ra$\,.\parn
First of all, let us consider the expression \eqref{T00II} for $\la 0|\Tiue_{00}(\q)|0\ra$
and recall once more the regularity properties of the various terms appearing therein
(see items (i)(ii) at the end of Section \ref{secAC}). In addition, let us notice
that the following asymptotic expansions hold for $\u \to 0$
(see \S 5 of \cite{NIST}; here and in the following $\gEM$ indicates the Euler-Mascheroni constant):

\begin{equation}\begin{array}{c}
\dd{\Ga\!\l({\u \over 2}-2\r) \,=\, {1 \over \u} + {1 \over 2} \l({3 \over 2} - \gEM\r) +\, O(\u) ~;} \vspace{0.2cm}\\
\dd{\Ga\!\l({\u + 1\over 2}\r) \,=\, \sqrt{\pi} - \sqrt{\pi} \l(\log 2 + {\gEM \over 2}\r) \u\, + \,O(\u^2) ~;}
\vspace{0.2cm} \\
\dd{\l({\mm \over \ee}\r)^{\!\u} =\, 1 + \u\,\log\!\l({\mm \over \ee}\r) + \, O(\u^2) ~.}
\end{array}\end{equation}

\noindent
Keeping in mind all these facts, by simple computations we obtain from Eq. \eqref{T00II}
({\footnote{Taking the regular part, as indicated in Eq. \eqref{T00ep}, amounts to remove
from the Laurent expansion for $\la 0 | \Tiue_{0 0}(\q) | 0 \ra$ at $\u = 0$
the pole term

$$ {\ee^4\,(1-8\xi) \over 32\pi^2}\;{1 \over \u} ~. $$

\noindent
This is the same divergent contribution appearing in the computation of the
renormalized energy density VEV for a scalar field of mass $\ee > 0$ in empty
space (with no external potentials or confining boundaries;
in this case, $\la 0 | \Tiue_{0 0}(\q) | 0 \ra$ is just given by the first
line of Eq. \eqref{T00II}).}})

\begin{equation}\begin{array}{c}
\dd{RP\Big|_{\u = 0}\, \la 0 | \Tiue_{0 0}(\q) | 0 \ra ~=~
{\ee^4 \over 8 \pi^2} \l[\,\l({5 \over 16} - \xi\r) + \l({1 \over 4} - 2\xi\r)\log\!\l({2\mm \over \ee}\r)\,\r]\, +}
\vspace{0.15cm} \\
\dd{+\; {1 \over 8 \pi^2\,r^4}\,
\Bigg[\l({1 \over 4} - \xi\r) \KM_{2}(2\,\ee\,r)\,
+ \l({1 \over 2} - 4\xi\r) \KM_{1}(2\,\ee\,r)\; +} \vspace{0.15cm}\\
\dd{+ \,\big(1 - 4\xi\big)\, \KM_{0}(2\,\ee\,r)
-\,{2r \over \at} \int_0^{+\infty}\!\! d\vv\; {e^{- {2r \over \at}\,\vv} \over (\vv + 1)^{2}}\;
\Bigg(\!\big(1 - 4 \xi\big)\, (\vv + 1)^2\,\KM_{0}\big(2\,\ee\,r\,(\vv + 1)\big)\;+ } \\
\dd{+ \l(\!(1 - 4 \xi)\,(\vv + 1) - {1 \over 2} \r) \KM_{1}\big(2\,\ee\,r\,(\vv\!+\!1)\big)
+ \l({1 \over 4} - \xi\r)\KM_{2}\big(2\,\ee\,r\,(\vv + 1)\big)\!\Bigg) \Bigg] ~.}
\label{T00ep}
\end{array}\end{equation}

\noindent
In the first line of Eq. \eqref{T00ep}, let us note the mass parameter $\mm$ which
has been introduced to regularize the field operator (see Eq. \eqref{Fiu}).
In the upcoming Section \ref{secIR} we will send to zero the infrared cutoff parameter $\ee$;
in view of this development, it is worthwhile to use the elementary identity

\begin{equation}
{2 r \over \at} \int_0^{+\infty}\!\! d\vv\; e^{-{2 r \over \at}\,\vv} \,=\, 1
\end{equation}

\noindent
in order to re-write the third line of Eq. \eqref{T00ep}. In this way we obtain the
following, equivalent version of the cited equation:

\begin{equation}\begin{array}{c}
\dd{RP\Big|_{\u = 0}\, \la 0 | \Tiue_{0 0}(\q) | 0 \ra ~=~
{\ee^4 \over 8 \pi^2} \l[\,\l({5 \over 16} - \xi\r) + \l({1 \over 4} - 2\xi\r)\log\!\l({2\mm \over \ee}\r)\,\r]\, +}
\vspace{0.15cm}\\
\dd{+\; {1 \over 8 \pi^2\,r^4}\,
\Bigg[\l({1 \over 4} - \xi\r) \KM_{2}(2\,\ee\,r)\,
+ \l({1 \over 2} - 4\xi\r) \KM_{1}(2\,\ee\,r)\; +} \vspace{0.15cm}\\
\dd{+ \; {2 r \over \at}\int_0^{+\infty}\! d\vv\; {e^{-{2r \over \at}\,\vv} \over (\vv+1)^2}\,
\Bigg(\big(1 - 4\xi\big)\, (\vv+1)^2\, \Big(\KM_{0}(2\,\ee\,r)- \KM_{0}\big(2\,\ee\,r\,(\vv + 1)\big)\Big) \;+ } \vspace{0.2cm}\\
\dd{ - \l( (1 - 4 \xi)\, (\vv + 1) - {1 \over 2} \r) \KM_{1}\big(2\,\ee\,r\,(\vv + 1)\big)
- \l({1 \over 4} - \xi\r) \KM_{2}\big(2\,\ee\,r\,(\vv + 1)\big) \Bigg)\, \Bigg] ~.}
\label{T00ep1}
\end{array}\end{equation}

\noindent
Similar results can be derived for the regular parts at $\u = 0$
of the other components of the regularized stress-energy VEV.
As shown in the next two sections, the dependence on $\mm$ disappears from all components
in the limit $\ee \to 0^+$. \parn

\section{Removal of the infrared cutoff: the limit \boldmath{$\ee \to 0^+$}}\label{secIR}
We already pointed out that the expressions derived in the previous section
for the regular part at $\u = 0$ of the regularized stress-energy VEV do still
depend on the infrared cutoff parameter $\ee \in (0,+\infty)$.
In this section we compute the limit $\ee \to 0^+$ of the above cited expressions;
in accordance with the general definition \eqref{Tren} of Section \ref{secgen},
this determines the renormalized VEV of the stress-energy tensor.
As usual, we illustrate for example the computation of the limit $\ee \to 0^+$
for $RP\big|_{\u = 0}\la 0|\Tiue_{00}(\q)|0\ra$, ultimately yielding the renormalized
energy density $\la 0|\Ti_{00}(\q)|0\ra_{ren}$. \parn
To this purpose, let us first consider the expression \eqref{T00ep1} for
$RP\big|_{\u = 0}\la 0|\Tiue_{00}(\q)|0\ra$.
Recalling the asymptotic behavior of the Bessel function $K_\si$ near zero
(see, e.g., Eq.s 10.30.2, 10.30.3 on page 252 of \cite{NIST}), it is easy to prove
that the function $\KM_{\si}$ defined in Eq. \eqref{KMdef} fulfills the following relations
(recall that $\gEM$ is the Euler-Mascheroni constant):

\begin{equation}
\lim_{\zz \to 0^+} \KM_\si(\zz) \,=\, 2^{\si - 1}\,\Ga(\si) \qquad
\mbox{for all $\si \in \C$ with $\Re\si > 0$} ~; \label{KM01}
\end{equation}

\begin{equation}
\KM_0(\zz) \,=\, - \log(\zz/2) + \gEM + O\big(\zz^2 \log \zz\big) \qquad \mbox{for $\zz \to 0^+$} ~.
\label{KM02}
\end{equation}

\noindent
In particular, let us remark that Eq. \eqref{KM01} gives

\begin{equation}
\lim_{\ee \to 0^+} \KM_{1}(2\,\ee\,r) \,=\, 1 ~, \quad
\lim_{\ee \to 0^+} \KM_{1}(2\,\ee\,r) \,=\, 2 \qquad \mbox{for all $r > 0$}~;
\end{equation}

\noindent
on the other hand, making reference to the expression in the third line of Eq. \eqref{T00ep1},
we can use Eq. \eqref{KM02} to infer that

\begin{equation}
\lim_{\ee \to 0^+} \Big[\KM_{0}(2\,\ee\,r)- \KM_{0}\big(2\,\ee\,r\,(\vv + 1)\big)\Big] \,=\,
\log(\vv + 1) \qquad \mbox{for all $r,\vv > 0$} ~.
\end{equation}

\noindent
In addition, let us point out that by Lebesgue's dominated convergence theorem the limit
$\ee \to 0^+$ can be evaluated before performing the integrations over $\vv$ in Eq. \eqref{T00ep1}. \parn
Summing up, the above arguments allow us to derive the following explicit expression
for the renormalized VEV of the energy density:\vspace{0.1cm} \parn

\vbox{

\begin{equation}
\la 0|\Ti_{00}(\q)|0\ra_{ren} ~=~ \label{T00ren}
\end{equation}

$$ {1 \over 8 \pi^2\,r^4}\, \Bigg[(1-6\xi) +
{2 r \over \at}\!\int_0^{+\infty}\!\! d\vv\; {e^{-{2r \over \at}\,\vv} \over (\vv+1)^2}\,
\Bigg(\big(1 - 4\xi\big)\,(\vv+1)^2 \log(\vv+1)- (1 - 4 \xi)\, (\vv + 1) + 2\xi \Bigg)\Bigg]\,. $$

}

\parn \noindent
To go on, it is useful to notice that the integral over $\vv \in (0,+\infty)$ appearing in Eq. \eqref{T00ren}
can be re-expressed in terms of the exponential integral function $\Eu$ (see, e.g., Chapter 6 of \cite{NIST}). \\
To be precise, let us introduce the function

\begin{equation}
\EE : (0,+\infty) \to \R\,, \qquad \rr \mapsto \EE(\rr) \,:=\, e^\rr\;\Eu(\rr) ~. \label{defEE}
\end{equation}

\noindent
Then, using a well-known integral representation for $\Eu$ (see, e.g., Eq. 6.2.2 on page 150 of \cite{NIST}),
by a simple change of the integration variable we obtain

\begin{equation}
\EE(\rr) \,= \int_0^{+\infty}\!\!d\vv\;{e^{-\rr\, \vv} \over \vv + 1} \qquad \mbox{for all $\rr > 0$} ~. \label{EEint}
\end{equation}

\noindent
Moreover, keeping in mind the definition \eqref{defEE} of $\EE$, by suitable integrations
by parts of the integral in the right-hand side of the above relation \eqref{EEint} we can
also prove the identities reported hereafter, for $\rr > 0$:

\begin{equation}
\int_0^{+\infty}\!\!d\vv\; e^{-\rr\,\vv}\, \log(\vv+1) \,=\, {1 \over \rr}\;\EE(\rr) ~, \qquad
\int_0^{+\infty}\!\!d\vv\; {e^{-\rr\,\vv} \over (\vv + 1)^2} \,=\, 1 - \rr\,\EE(\rr) ~.
\end{equation}

\noindent
We can use the results mentioned above to re-express Eq. \eqref{T00ren} as

\begin{equation}
\la 0|\Ti_{00}(\q)|0\ra_{ren} \,=\,
{1 \over 8 \pi^2\,r^4}\, \Bigg[(1-6\xi) + 2\xi\,\rr
+ \Big(\big(1 - 4\xi\big)\;(1-\rr) - 2\xi\,\rr^2 \Big)\,\EE(\rr)\Bigg]_{\rr \,=\, {2r/\at}}\,.
\label{T00ren2}
\end{equation}

\noindent
Arguments analogous to those presented in this section can be employed to determine
all the other components of the renormalized stress-energy VEV $\la 0|\Ti_{\mu\nu}(\q)|0\ra_{ren}$.
In the upcoming, conclusive Section \ref{secTmn} we collect our final results for these quantities
and discuss their asymptotic behaviors in various regimes.

\section{The renormalized stress-energy VEV}\label{secTmn}
We now give the final form of our results separating the conformal and non-conformal parts
of the renormalized stress-energy VEV, according to the general scheme of Section \ref{secgen}
(see, especially, Eq. \eqref{Tcnc0} and related comments).
Using the spherical coordinates $\q = (r,\te,\vfi)$, we have the relation

\begin{equation}
\la 0|\Ti_{\mu\nu}(\q)|0\ra_{ren} \,=\, \Tc_{\mu\nu}(\q) + \big(\xi - \xi_c\big)\, \Tnc_{\mu\nu}(\q) ~,
\qquad \xi_c \,:=\, {1 \over 6} ~, \label{Tcnc}
\end{equation}

\noindent
defining the conformal and non-conformal parts $\Tc_{\mu\nu}$ and $\Tnc_{\mu\nu}$.
The non-zero components in this representation are as follows:

\begin{equation}\begin{array}{c}
\dd{\Tc_{00}(\q) \,=\, {1 \over 24 \pi^2\,r^4}\,
\Big[\rr + \big(1 - \rr - \rr^2 \big)\,\EE(\rr)\Big]_{\rr \,=\, {2r/\at}} ~,}\vspace{0.1cm}\\
\dd{\Tnc_{00}(\q) \,=\, -\;{1 \over 4 \pi^2\,r^4}\,
\Big[3 - \rr + \big(2 -  2 \rr + \rr^2 \big)\,\EE(\rr)\Big]_{\rr \,=\, {2r/\at}} ~;} \label{T00cnc}
\end{array}\end{equation}

\begin{equation}\begin{array}{c}
\dd{\Tc_{rr}(\q) \,=\, {1 \over 24 \pi^2\,r^4}\,
\Big[1 - \big(1+\rr\big)\, \EE(\rr) \Big]_{\rr \,=\,{2r/\at}} ~,} \vspace{0.1cm}\\
\dd{\Tnc_{rr}(\q) \,=\, -\;{1 \over 2 \pi^2\,r^4}\,
\Big[1 + \big(2-\rr\big)\, \EE(\rr)\Big]_{\rr\,=\,{2r/\at}} ~;}
\end{array}\end{equation}

\begin{equation}\begin{array}{c}
\dd{\Tc_{\te\te}(\q) \,=\, \Tc_{\vfi\vfi}(\q)/\sin^2\!\te \,=\, -\;{1 \over 48 \pi^2\,r^2}\,
\Big[\big(1-\rr\big)- \big(2-\rr^2\big)\,\EE(\rr)\Big]_{\rr\,=\,{2r/\at}} ~,} \vspace{0.1cm}\\
\dd{\Tnc_{\te\te}(\q) \,=\, \Tnc_{\vfi\vfi}(\q)/\sin^2\!\te \,=\, {1 \over 4 \pi^2\,r^2}\, \Big[\big(4 -\rr\big)
+ \big(4 - 3 \rr + \rr^2\big)\, \EE(\rr)\Big]_{\rr\,=\,{2r/\at}} ~.} \label{T22cnc} \vspace{0.15cm}
\end{array}\end{equation}

\begin{figure}[t!]
    \centering
        \begin{subfigure}[b]{0.45\textwidth}
                \includegraphics[width=\textwidth]{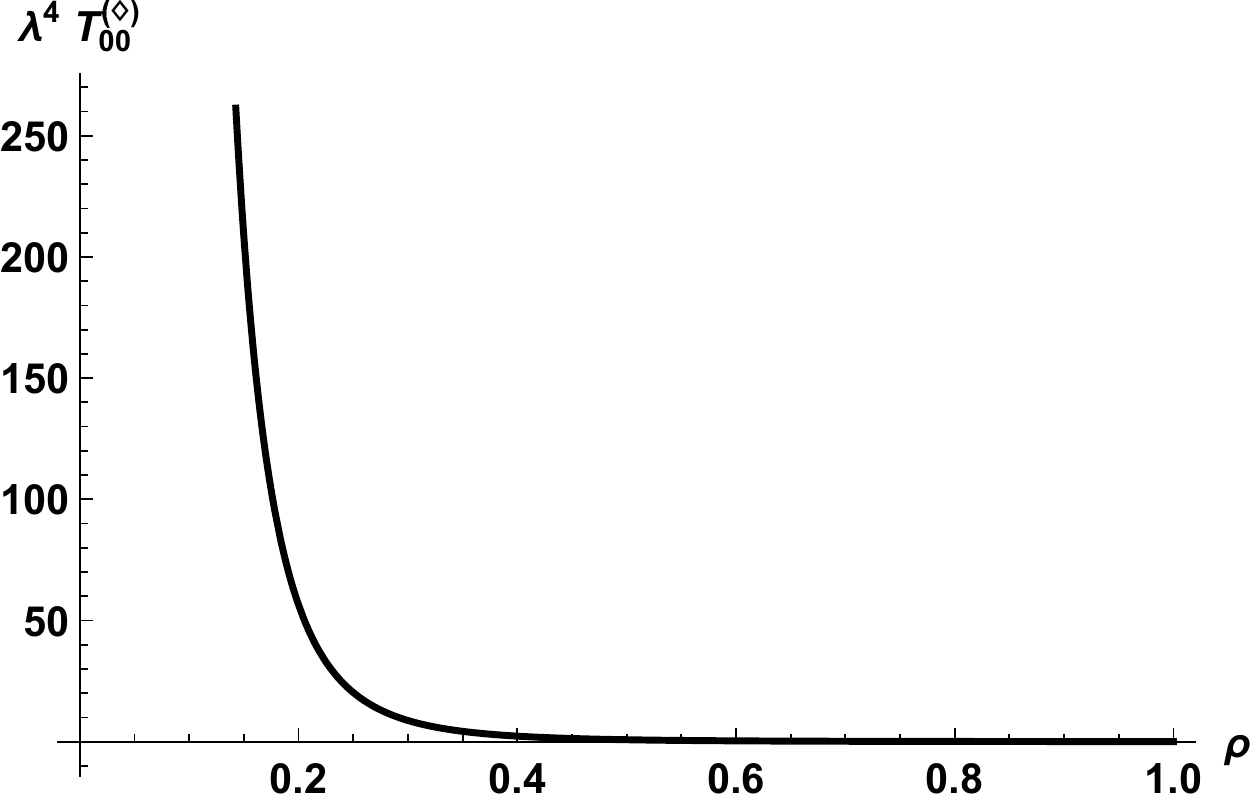}
        \end{subfigure}\hspace{1cm}
        \begin{subfigure}[b]{0.45\textwidth}
                \includegraphics[width=\textwidth]{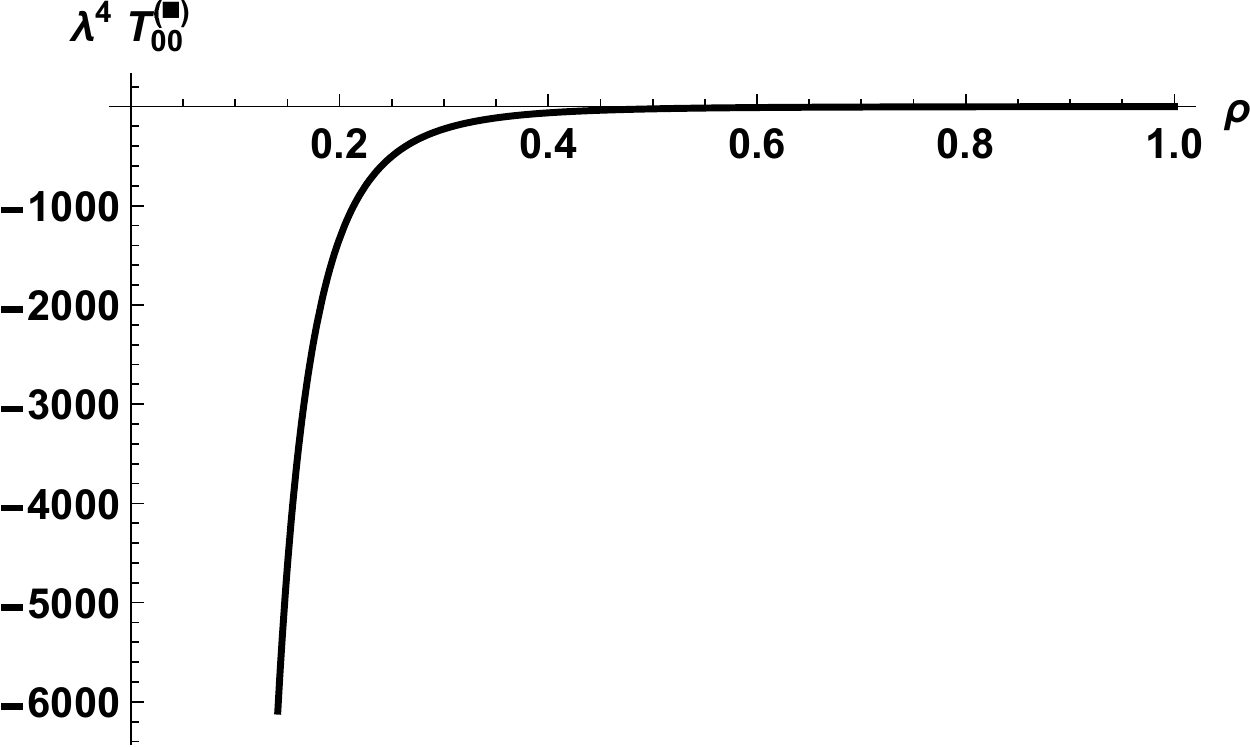}
        \end{subfigure}
        \caption{graphs of $\at^4\,\Tc_{00}$ and $\at^4\,\Tnc_{00}$ as functions of $\rr := 2r/\at$\,.}\label{fig:T00}
\end{figure}

\begin{figure}[t!]
    \centering
        \begin{subfigure}[b]{0.45\textwidth}
                \includegraphics[width=\textwidth]{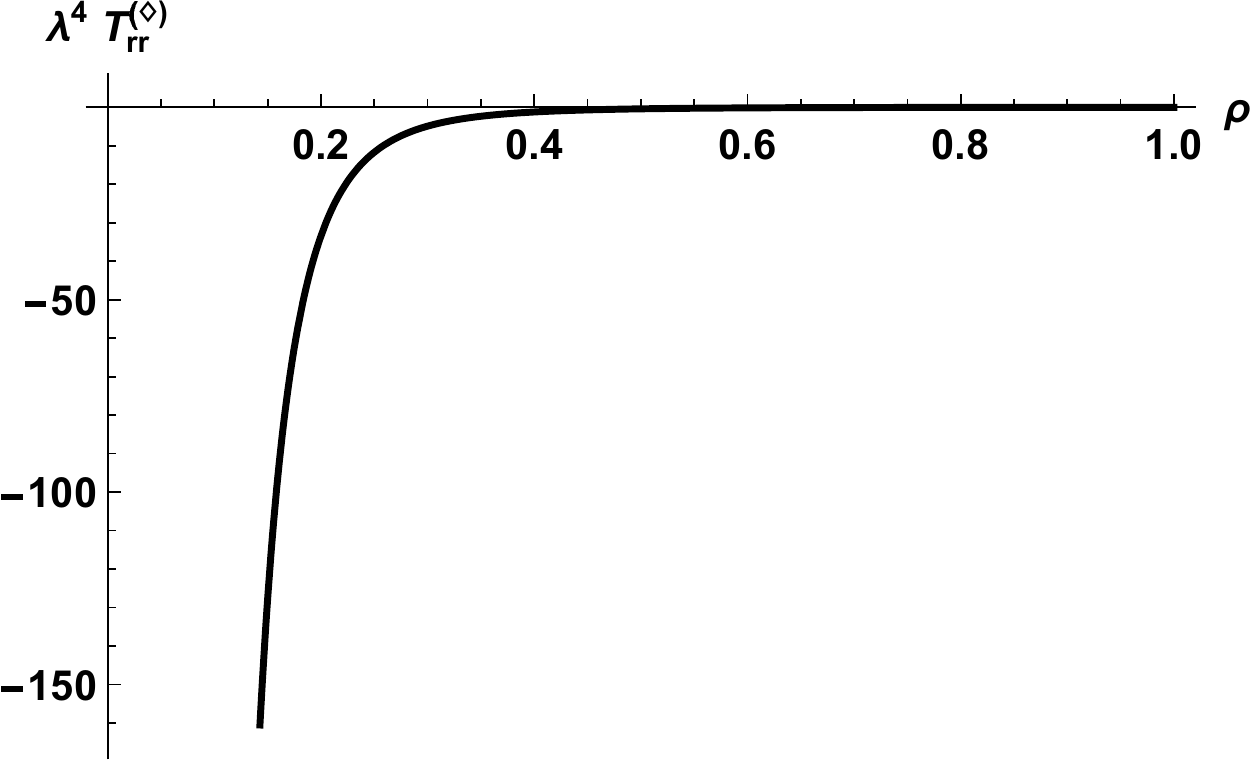}
        \end{subfigure}\hspace{1cm}
        \begin{subfigure}[b]{0.45\textwidth}
                \includegraphics[width=\textwidth]{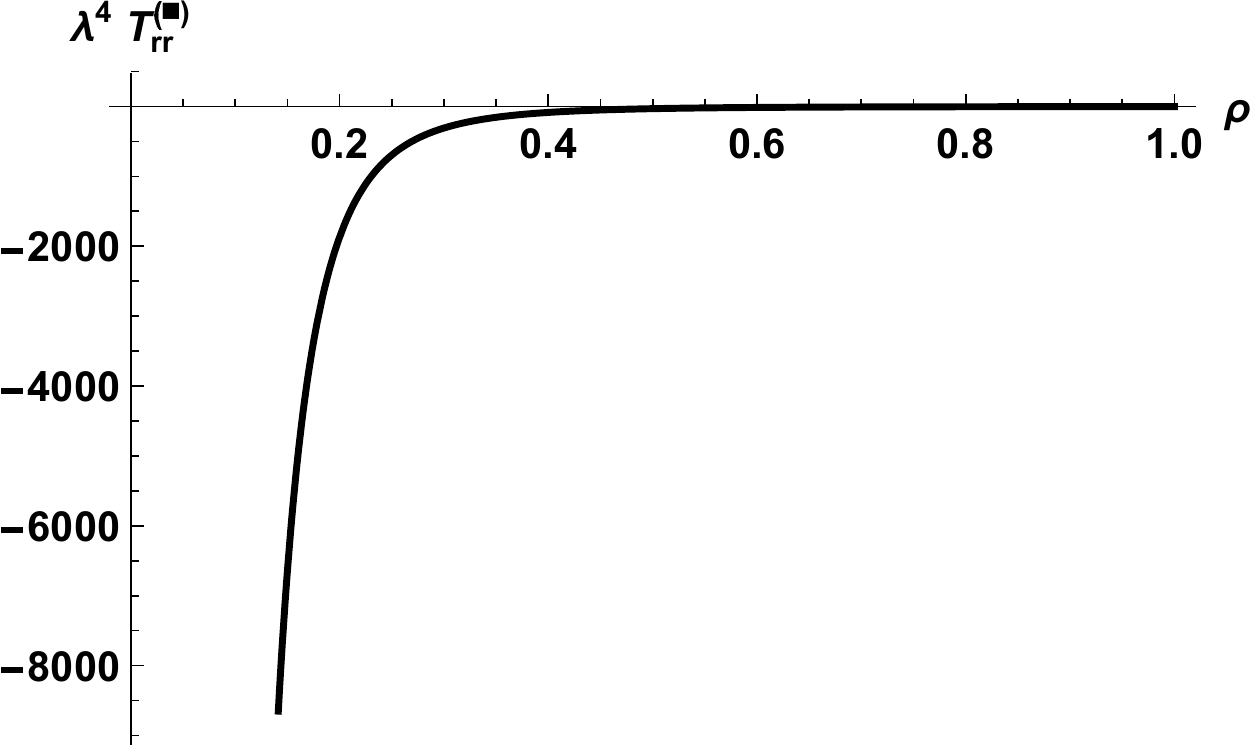}
        \end{subfigure}
        \caption{graphs of $\at^4\,\Tc_{rr}$ and $\at^4\,\Tnc_{rr}$ as functions of $\rr := 2r/\at$\,.}\label{fig:T11}
\end{figure}

\begin{figure}[t!]
    \centering
        \begin{subfigure}[b]{0.45\textwidth}
                \includegraphics[width=\textwidth]{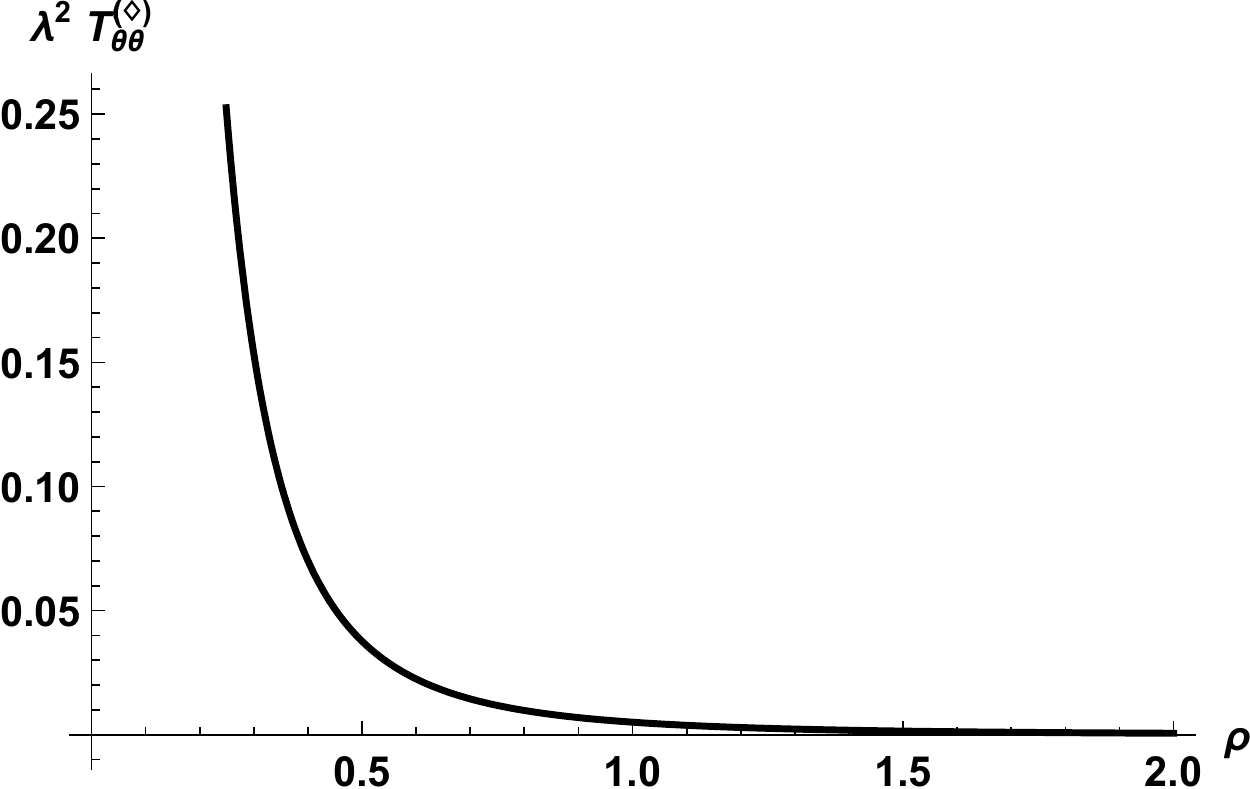}
        \end{subfigure}\hspace{1cm}
        \begin{subfigure}[b]{0.45\textwidth}
                \includegraphics[width=\textwidth]{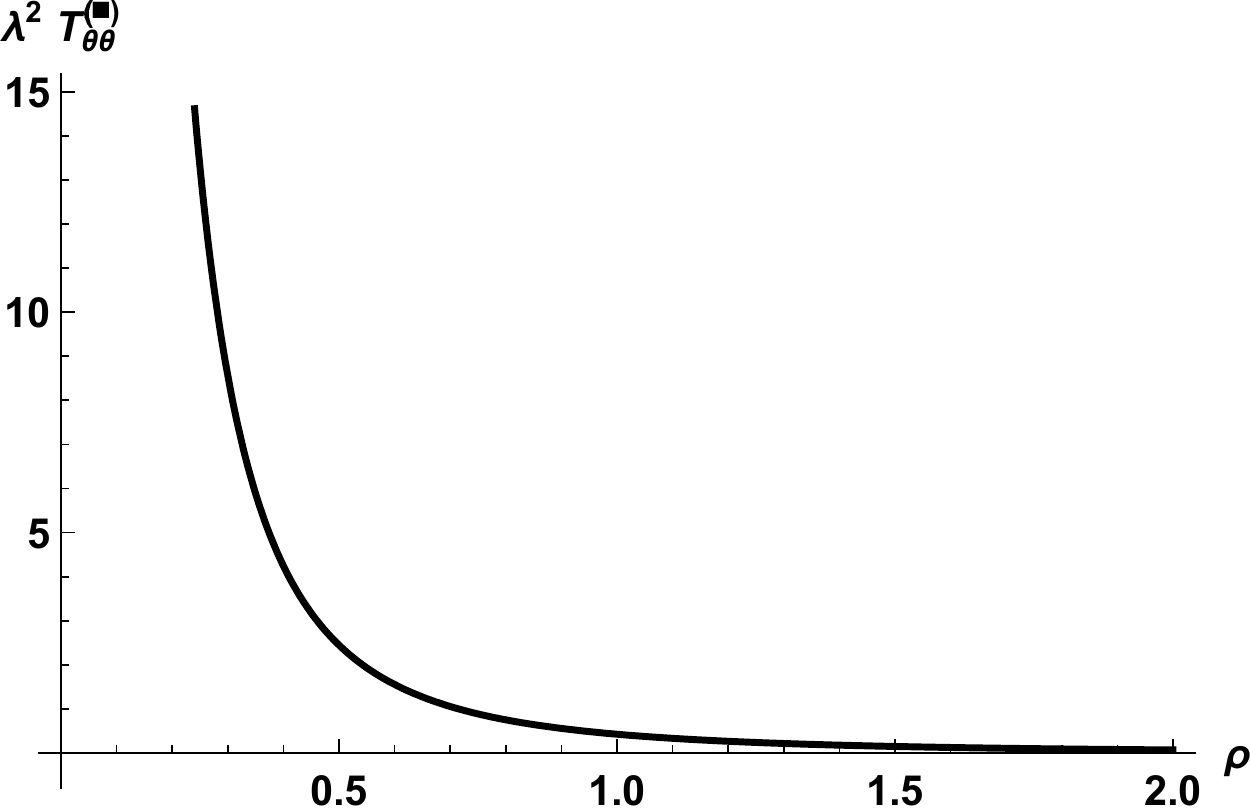}
        \end{subfigure}
        \caption{graphs of $\at^2\,\Tc_{\te\te}$ and $\at^2\,\Tnc_{\te\te}$ as functions of $\rr := 2r/\at$\,.}\label{fig:T22}
\end{figure}

\noindent
From the explicit expressions reported above, it is evident that $\at^4\,\Tc_{00}$, $\at^4\,\Tc_{rr}$
and $\at^2\,\Tc_{\te\te}$ as well as their non-conformal counterparts depend solely
on the dimensionless variable $\rr := 2r/\at$; the graphs of these functions of $\rr$
are reported in Figures \ref{fig:T00}-\ref{fig:T22}.
\captionsetup{justification=centering,margin=2cm}
\parn
In the following subsections \ref{subsmallr}, \ref{sublarger} we derive the asymptotic
expansions of $\Tc_{\mu\nu}(\q), \Tnc_{\mu\nu}(\q)$ when $\rr := 2r/\at$ tends to $0^+$
and to $+\infty$. These expansions have a twofold interpretation:
indeed, they determine the dominant contributions in the renormalized stress-energy VEV
for small and large values of the radial coordinate $r$ or, alternatively,
for large and small values of the parameter $\at$.

\subsection{Asymptotic expansions for $\rr = 2r/\at \to 0^+$}\label{subsmallr}
Let us consider the definition \eqref{defEE} of the map $\EE$, involving the
exponential integral function $\Eu$; using a well-known series representation for the
latter (see, e.g., Eq. 5.4.14 and Eq. 6.6.2 of \cite{NIST}), it is easily shown that

\begin{equation}
\EE(\rr) \,=\, - \sum_{n = 0}^{+\infty} \Big(\log \rr + \gEM - H_n\Big)\, {\rr^n \over n!}
\qquad \mbox{for all $\rr > 0$} \label{EEser}
\end{equation}

\noindent
(as usual, $\gEM$ is the Euler-Mascheroni constant; $H_n := \sum_{j = 1}^n 1/j$ is the
$n$-th harmonic number).\parn
Of course, the series representation \eqref{EEser} determines the asymptotic expansion
of $\EE(\rr)$ for $\rr \to 0^+$. In particular, this allows us to infer the following relations,
for $\rr = 2r/\at \to 0^+$:

\begin{equation}\begin{array}{c}
\dd{\Tc_{00}(\q) \,=\, -\,{1 \over 24\pi^2 r^4}\;
\Big[\log \rr + \gEM + O(\rr)\Big] ~,} \vspace{0.15cm}\\
\dd{\Tnc_{00}(\q) \,=\, {1 \over 2\pi^2 r^4}\;
\Big[\log \rr + \gEM - {3 \over 2} + O(\rr)\Big]~ ;} \vspace{0.cm}
\label{T00r0}
\end{array}\end{equation}

\begin{equation}\begin{array}{c}
\dd{\Tc_{rr}(\q) \,=\, {1 \over 24 \pi^2 r^4}\;
\Big[\log\rr + \gEM + 1 + O(\rr)\Big] ~,} \vspace{0.15cm}\\
\dd{\Tnc_{rr}(\q) \,=\, -\,{1 \over \pi^2 r^4}\;
\Big[\log\rr + \gEM - {1 \over 2} + O(\rr)\Big] ~;} \vspace{0.1cm}
\end{array}\end{equation}

\begin{equation}\begin{array}{c}
\dd{\Tc_{\te\te}(\q) \,=\, -\,{1 \over 24 \pi^2 r^2}\;
\Big[\log\rr + \gEM + {1 \over 2} + O(\rr)\Big] ~,} \vspace{0.15cm}\\
\dd{\Tnc_{\te\te}(\q) \,=\, -\,{1 \over \pi^2 r^2}\;
\Big[\log\rr + \gEM - 1 + O(\rr)\Big] ~.} \vspace{0.2cm}\\
\end{array}\end{equation}

\noindent
The above relations show that all the non-vanishing components of the renormalized
stress-energy VEV diverge near the origin $r = 0$,
where the point impurity is placed. In particular, Eq. \eqref{T00r0} makes patent the fact
that the renormalized energy density $\la 0|\Ti_{00}(\q)|0\ra_{ren}$ possesses a non-integrable
singularity at $r = 0$; in consequence of this, it is not possible to define the total energy
for the configuration under analysis simply by integration over $\Om = \R^3 \setminus \{\b0\}$
of $\la 0|\Ti_{00}(\q)|0\ra_{ren}$.
Here, we limit ourselves to mention that the appearance of problematic features of the above
kind is rather typical in Casimir-type computations. (See, e.g., \cite{FPBook}. In general,
the strategy to obtain the renormalized total energy VEV consists in exchanging the order
of the operations involved: one first integrates the regularized energy density and then
takes the regular part at $\u = 0$. \S 3.5 of the cited book contains some comments on this subject.)

\subsection{Asymptotic expansions for $\rr = 2r/\at \to +\infty$}\label{sublarger}
Recalling again the definition \eqref{defEE} of $\EE$ and using a known asymptotic
expansion of the exponential integral function $\Eu$ for large values of the argument
(see, e.g., Ex. 2.2 on page 112 of \cite{Olv}), for any $M \in \N$ we get

\begin{equation}
\EE(\rr) \,=\, {1 \over \rr} \sum_{m = 0}^{M} {(-1)^m\, m! \over \rr^m}\,
+ O\!\l({1 \over \rr^{M+2}}\r) \qquad \mbox{for $\rr \to +\infty$} ~.
\end{equation}

\noindent
The above result allows us to derive the following asymptotic relations, for $\rr = 2r/\at \to +\infty$:

\begin{equation}\begin{array}{c}
\dd{\Tc_{00}(\q) \,=\, {1 \over 8 \pi^2 r^4}
\l[\,{1 \over \rr^2} - {16 \over 3 \rr^3}  + {30 \over \rr^4} - {192 \over \rr^5} + O\!\l({1 \over \rr^6}\r)\r] ,} \vspace{0.15cm}\\
\dd{\Tnc_{00}(\q) \,=\, -\,{3 \over 2\pi^2 r^4}
\l[\,{1 \over \rr} - {2 \over \rr^2} + {20 \over 3\rr^3} - {28 \over \rr^4} + O\!\l({1 \over \rr^5}\r) \r] ;}
\label{T00inf}
\end{array}\end{equation}

\begin{equation}\begin{array}{c}
\dd{\Tc_{rr}(\q) \,=\, -\,{1 \over 24 \pi^2 r^4}
\l[\,{1 \over \rr^2} - {4 \over \rr^3} + {36 \over \rr^4} - {96 \over \rr^5} + O\!\l({1 \over \rr^6}\r)\r],}\vspace{0.15cm}\\
\dd{\Tnc_{rr}(\q) \,=\, -\,{3 \over 2\pi^2 r^4}
\l[\,{1 \over \rr} - {4 \over 3\rr^2} + {10 \over 3 \rr^3}- {12 \over \rr^4} + O\!\l({1 \over \rr^5}\r) \r] ;} \vspace{0.15cm}
\end{array}\end{equation}

\begin{equation}\begin{array}{c}
\dd{\Tc_{\te\te}(\q) \,=\, {1 \over 12 \pi^2 r^2}
\l[\,{1 \over \rr^2} - {5 \over \rr^3} + {27 \over \rr^4} - {168 \over \rr^5} + O\!\l({1 \over \rr^6}\r)\r],}\vspace{0.15cm}\\
\dd{\Tnc_{\te\te}(\q) \,=\, {9 \over 4\pi^2 r^2}
\l[\,{1 \over \rr} - {16 \over 9 \rr^2} + {50 \over 9\rr^3} - {24 \over \rr^4} + O\!\l({1 \over \rr^5}\r)\r].}
\label{T22inf}
\end{array}\end{equation}

\noindent
The above asymptotic expansions show that the renormalized
stress-energy VEV vanishes quite rapidly for large values of $r$, that is for large distances
from the impurity. \\
Apart from that, Eq.s \eqref{T00inf}-\eqref{T22inf} also allow us to infer that

\begin{equation}
\lim_{\at \to 0^+} \la 0|\Ti_{\mu\nu}(\q)|0\ra_{ren} \,=\, 0 ~. \label{limzer}
\end{equation}

\noindent
We recall that, for $\at = 0$, the quantum field theory under analysis reduces to that
of a free scalar field in empty Minkowski spacetime; in this regard, the identity
\eqref{limzer} matches the physically sensible fact that the renormalized VEV of the
stress-energy tensor vanishes identically when no potential (or no boundary) is present.

\vspace{6pt}

\acknowledgments{This work was partly supported by: INdAM, Gruppo Nazionale per la Fisica Matematica;
INFN; MIUR, PRIN 2010 Research Project “Geometric and analytic theory of Hamiltonian systems in finite
and infinite dimensions.”; Universit\`{a} degli Studi di Milano.}



\abbreviations{The following abbreviations are used in this manuscript:\\

\noindent
\begin{tabular}{@{}ll}
VEV & vacuum expectation value
\end{tabular}}


\reftitle{References}


\end{document}